Technical Report

# Open Mobile API:
# Accessing the UICC on Android Devices

Michael Roland

University of Applied Sciences Upper Austria
Josef Ressel Center u'smile
michael.roland@fh-hagenberg.at

Michael Hölzl

Johannes Kepler University Linz
Institute of Networks and Security
hoelzl@ins.jku.at

**Abstract** This report gives an overview of secure element integration into Android devices. It focuses on the Open Mobile API as an open interface to access secure elements from Android applications. The overall architecture of the Open Mobile API is described and current Android devices are analyzed with regard to the availability of this API. Moreover, this report summarizes our efforts of reverse engineering the stock ROM of a Samsung Galaxy S3 in order to analyze the integration of the Open Mobile API and the interface that is used to perform APDU-based communication with the UICC (Universal Integrated Circuit Card). It further provides a detailed explanation on how to integrate this functionality into CyanogenMod (an after-market firmware for Android devices).

This work has been carried out within the scope of "u'smile", the Josef Ressel Center for User-Friendly Secure Mobile Environments, funded by the Christian Doppler Gesellschaft, A1 Telekom Austria AG, Drei-Banken-EDV GmbH, LG Nexera Business Solutions AG, NXP Semiconductors Austria GmbH, and Österreichische Staatsdruckerei GmbH in cooperation with the Institute of Networks and Security at the Johannes Kepler University Linz. Moreover, this work has been carried out in close cooperation with the project "High Speed RFID" within the EU programme "Regionale Wettbewerbsfähigkeit OÖ 2007–2013 (Regio 13)" funded by the European Regional Development Fund (ERDF) and the Province of Upper Austria (Land Oberösterreich).

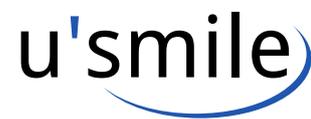

Revision 1.0
January 11, 2016

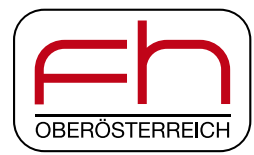

# Contents

# 1. Introduction

The Open Mobile API [18] is created and maintained by SIMalliance, a non-profit trade association that aims at creating secure, open and interoperable mobile services. It defines a platform-independent architecture for interfacing secure elements on mobile devices and specifies a programming language independent API for integrating secure element access into mobile applications. Using the Open Mobile API, mobile applications can interact with secure elements of virtually any form-factor integrated in mobile devices, e.g. an embedded secure element, a universal integrated circuit card (UICC), or an advanced security (micro) SD card (ASSD). Consequently, developers can make use of enhanced security capabilities provided by such secure elements.

The project "Secure Element Evaluation Kit for the Android platform" (SEEK-for-Android, [4]) provides the "Smartcard API" as an open-source implementation of the Open Mobile API specification for the Android operating system platform. The project releases patches to integrate the Smartcard API into the Android Open Source Platform (AOSP). These patches include the smartcard subsystem (consisting of the smartcard system service and the smartcard API framework) and interface modules to access different forms of secure elements (an ASSD, an embedded secure element that is accessible through Android's proprietary secure element API, and a UICC accessible through the radio interface layer (RIL) with standardized AT commands [3, 7]).

Meanwhile, many smartphones (in particular those by Samsung and Sony) ship with an implementation of the Open Mobile API in their stock ROM. These implementations give access to a UICC-based secure element and (on some devices) also to an embedded secure element. The vendor-specific implementations look similar to the SEEK implementation and differ only slightly in their behavior (e.g. access control mechanisms). Due to the standardized API definition, apps compiled against the SEEK smartcard API framework seamlessly integrate with the vendor-specific implementations shipped in stock ROMs.

Unfortunately, there is a gap between the open-source implementation provided by SEEK and the vendor-specific implementations provided in stock ROMs: the secure element interface modules. Particularly interfacing the UICC through the baseband modem is not well standardized. The baseband modem is accessed through the radio interface layer (RIL) which comprises of a vendor-specific low-level library, a RIL system daemon, and the telephony framework. The RIL is a device-specific closed-source component and varies between hardware-platforms. As a consequence, the SEEK implementation of the UICC interface module is not compatible with these proprietary interfaces and, therefore, is not usable with devices in the field. While this does not pose a problem when the smartcard API is used with the stock ROM shipped by the device manufacturer, it effectively prevents access to the UICC on custom ROMs and alternative Android distributions (like CyanogenMod[1]).

This is exactly what happened to us on the Samsung Galaxy S3. The stock ROM supports access to the UICC—and, after properly configuring the access control policies on our SIM cards, we could easily access them through the smartcard API. However, as soon as we switched over to CyanogenMod (or our custom "SuperSmile" ROM[2], which is based on SuperNexus and borrows the RIL implementations from the CyanogenMod project) we were no longer able to access the UICC. Even though we added the whole SEEK implementation including the interface module for UICC-access to our custom build, the proprietary RIL implementation on the Samsung Galaxy S3 does not provide the necessary extensions that are expected by the SEEK implementation.

As we needed access to a UICC-based secure element on our customizable ROM, we decided to analyze and reverse-engineer the implementation that is integrated into Samsung's stock ROM in order to build our own UICC interface module.

This report gives an overview of secure element integration into Android devices. It provides a deep insight of how secure element hardware is embedded into smartphones and how these secure elements can be accessed by applications on current Android devices. The focus of this report is on the Open Mobile API as an open middleware layer to access secure elements. We describe the overall architecture of the Open Mobile API and how this architecture is integrated into Android devices. Current Android devices are analyzed with regard to the availability of the Open Mobile API. Finally, we summarize our efforts of reverse engineering the stock ROM of a Samsung Galaxy S3 in order to evaluate the integration of the Open Mobile API in an existing device. We analyze the interface that is used to perform APDU-based communication with the UICC and provide a detailed explanation on how to integrate this functionality into CyanogenMod in order to enable UICC-access in a custom ROM.

---

[1] http://www.cyanogenmod.org/
[2] SuperSmile ROM: https://usmile.at/downloads

## 2.  Secure Element Integration

A secure element (SE) is a secure, tamper-resistant smartcard microchip that is integrated into a mobile device. In NFC card emulation mode, the SE is used to emulate a contactless smartcard over the RF front-end of the NFC controller. The NFC controller routes all[3] communication to the secure element in that case. Moreover, an SE can be accessed by apps running on the main application processor. Hence, apps can take advantage of security features and applications running on the secure element.

A secure element can be a dedicated microchip that is embedded into the mobile device hardware (embedded SE). Another possibility is the combination of the secure element functionality with another smartcard/security device that is used within the mobile device. For instance, a UICC (also known as the subscriber identity module, SIM card) is a smartcard that is already present in many mobile devices (particularly in mobile phones). Another security device that is available equipped with smartcard technology is micro SD (secure digital) cards.

Many secure elements (e.g. NXP's SmartMX) are standard smartcard ICs as used for regular contact and contactless smartcards. They share the same hardware and software platforms. The only difference is the interface they provide: Instead of (or in addition to) a classic smartcard interface according to ISO/IEC 7816 (for contact cards) or an antenna with an interface according to ISO/IEC 14443 (for contactless cards), the secure element has a direct interface for the connection to the NFC controller. Such interfaces are, for instance, the NFC Wired Interface (NFC-WI, [1]) and the Single Wire Protocol (SWP, [2]).

### 2.1  Embedded Secure Element

Various NFC-enabled mobile devices ship with an embedded secure element that is soldered into the mobile device hardware. Sometimes an embedded SE is combined into a single package with the NFC controller. An example for such a combined chip module is NXP's PN65N which contains a PN544 NFC controller and a secure element from NXP's SmartMX series.

Figure 1 shows the interconnection of the components in a mobile device with embedded SE on both a physical and a logical layer. An embedded SE is usually only wired to the NFC controller and it uses the NFC controller as its gateway to the outside world. Typical interfaces for connecting an embedded SE to the NFC controller are [17]

---

[3]Current NFC controllers are capable of dynamically routing card emulation communication to multiple secure elements and the host controller (main application processor) based on a routing table (cf. [11]). The routing decision can be based on application identifiers and on RF protocol types and parameters.

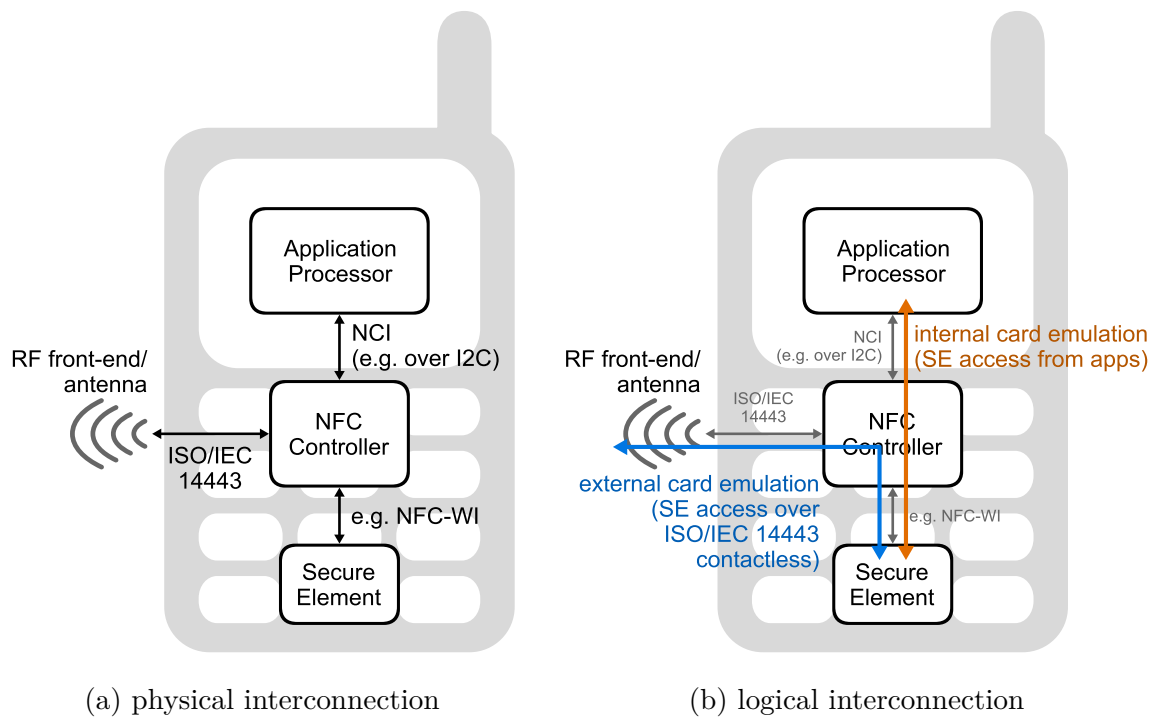

(a) physical interconnection      (b) logical interconnection

Figure 1: Embedded secure element in a mobile device

- NFC Wired Interface (NFC-WI),

- S2C (NXP's proprietary predecessor of NFC-WI),

- Single Wire Protocol (SWP),

- Digital Contact Less Bridge (DCLB),

- ISO/IEC 7816 (standard smartcard interface),

- Serial Peripheral Interface (SPI), and

- Inter-Integrated Circuit Bus ($I^2C$).

In external card emulation mode, the secure element is accessed over the contactless interface (ISO/IEC 14443) of the NFC controller. The NFC controller acts as the RF modem and routes communication from the RF antenna to the secure element chip. In internal card emulation mode, apps running on the main application processor access the secure element by using the NFC controller as a gateway that forwards communication to the SE. The NFC controller typically tags the communication so that the SE can distinguish between external and internal card emulation mode.

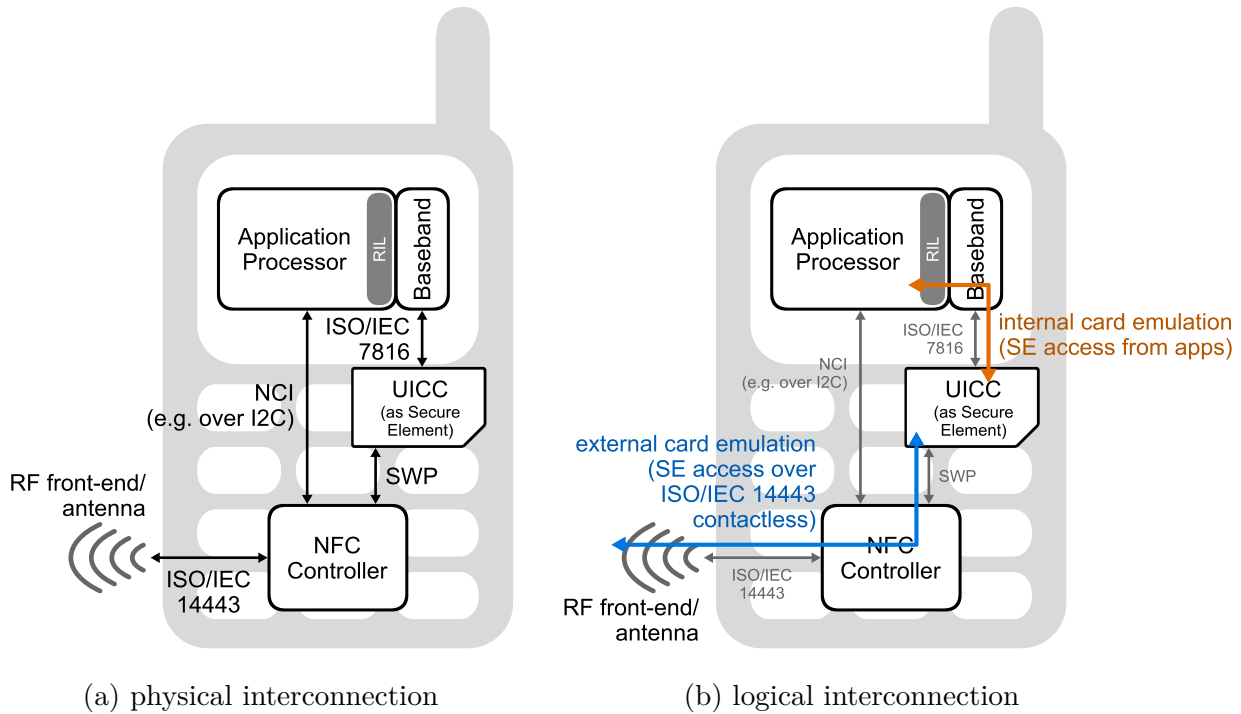

(a) physical interconnection          (b) logical interconnection

Figure 2: UICC-based secure element in a mobile device

## 2.2  Universal Integrated Circuit Card (UICC)

Many mobile devices (particularly smartphones) have support for a pluggable universal integrated circuit card (UICC) that provides the identity for access to mobile networks. In an NFC-enabled mobile device, a special UICC with support for single wire protocol (SWP) and support for application loading in the field can be used as an NFC secure element.

Figure 2 shows the interconnection of the components in a mobile device with UICC-based SE on both a physical and a logical layer. As with any other device that contains a UICC (or SIM card), the UICC is connected to the baseband processor over its ISO/IEC 7816 contact smartcard interface. In addition, an NFC-enabled UICC is directly connected to the NFC controller through the SWP interface.

In external card emulation mode, the UICC-based secure element is accessed over the contactless interface (ISO/IEC 14443) of the NFC controller. The NFC controller acts as the RF modem and routes communication from the RF antenna over the SWP connection to the UICC. In internal card emulation mode, apps running on the main application processor access the UICC-based secure element through the radio interface layer (RIL) by using the baseband processor as a gateway that forwards communication to the UICC over the ISO/IEC 7816 contact smartcard interface.

As the NFC controller is not used for internal card emulation using the UICC, this scenario is also viable on devices without an NFC controller. Hence, apps running on the application processor could still take advantage of the UICC as secure storage and trusted execution environment even if a device does not support NFC. In that case, the UICC is only connected to the baseband processor (see Fig. 3). Consequently, the UICC does not need have an SWP interface. The only requirement is that the radio interface layer (RIL) and the baseband provide APDU (application protocol data unit) based access to the UICC.

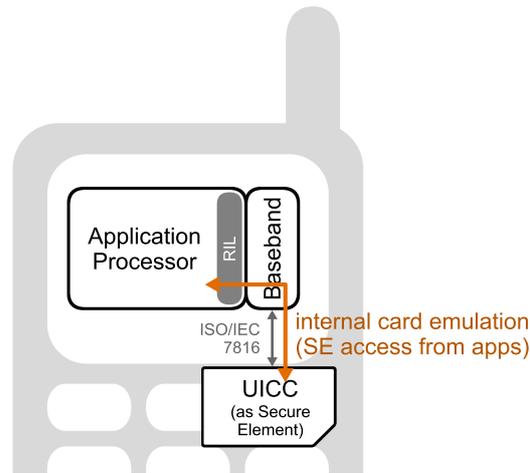

Figure 3: UICC-based secure element in a mobile device without NFC

## 2.3 Micro SD Card (smartSD/ASSD)

The standardized way of integrating an SD card based secure element are smartSD memory cards [16]. These micro SD cards permit APDU-based access from the application processor to the smartcard functionality in internal card emulation mode through the Advanced Security Secure Digital (ASSD, [14]) extension to the SD card interface and/or through a proprietary interface based on file-system I/O commands.

For external card emulation, a smartSD card may share the contactless front-end with the NFC controller—comparable to the UICC scenario—over single wire protocol in an NFC-enabled mobile device (see Fig. 4). An addendum [15] to the SD specification defines how a micro SD card can expose the necessary pin for the SWP interface. However, this only works if the mobile device has an NFC controller, and if the SD card slot of the mobile device also connects that SWP pin to the NFC controller, which is not the case in most mobile devices.

Instead of support for SWP, some smartSD cards contain their own contactless front-end (see Fig. 5). This permits integration of the smartSD card into virtually any device. Depending on the card, the antenna can be either embedded directly into the micro SD card (and possibly extended through a passive antenna booster) or it can be embedded into the mobile device and attached to dedicated pins of the SD card. Using such a card that exposes its own RF interface in an NFC-enabled device usually leads to the problem that the smartSD card may be discovered by the reader/writer capabilities of the NFC device. This depends on the position of the NFC antennas and may even only happen when the NFC reader antenna and the antenna of the smartSD card are inductively coupled to each other by an externally applied resonant circuit (e.g. an NFC tag). Consequently, such cards should not be used in NFC-enabled mobile devices.

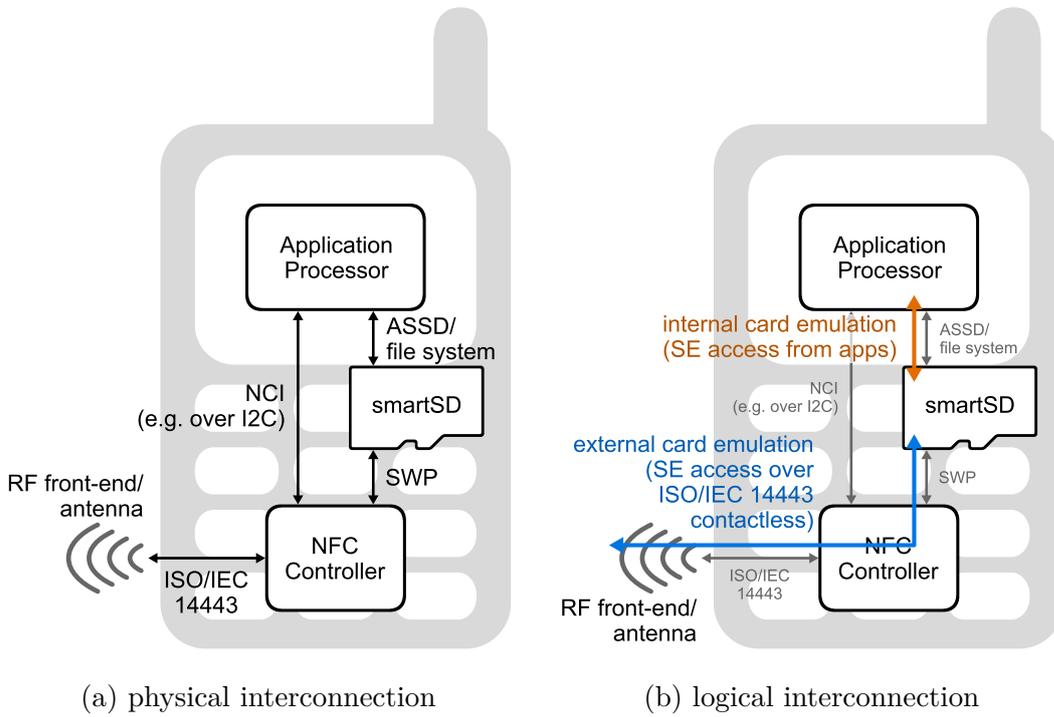

(a) physical interconnection          (b) logical interconnection

Figure 4: smartSD card with SWP support in a mobile device

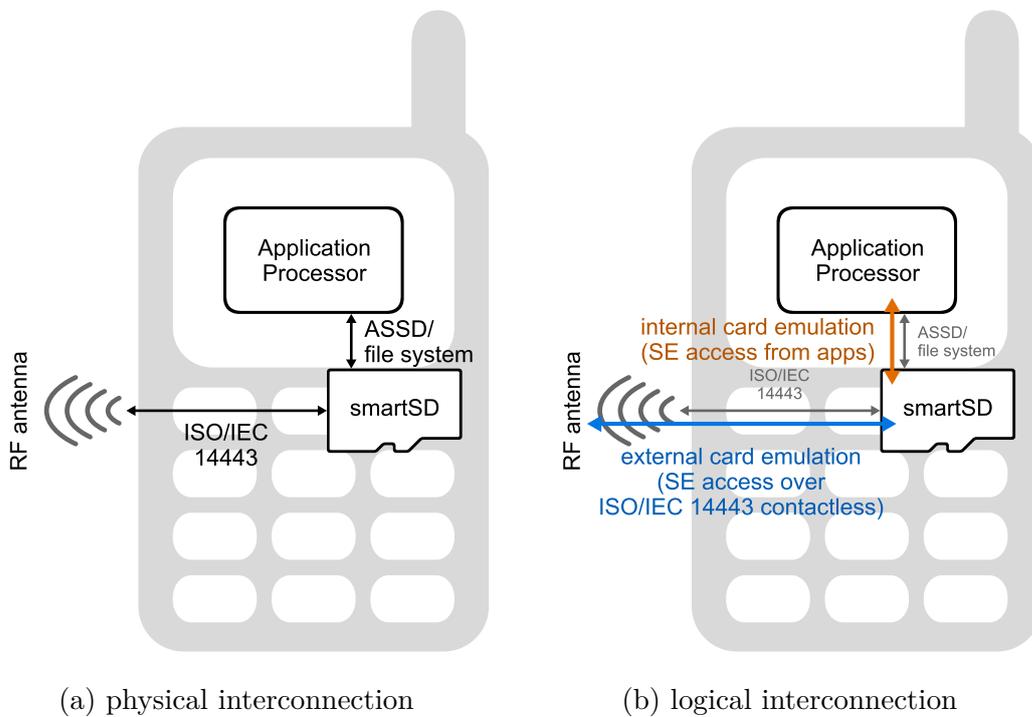

(a) physical interconnection          (b) logical interconnection

Figure 5: smartSD card with direct RF interface in a mobile device

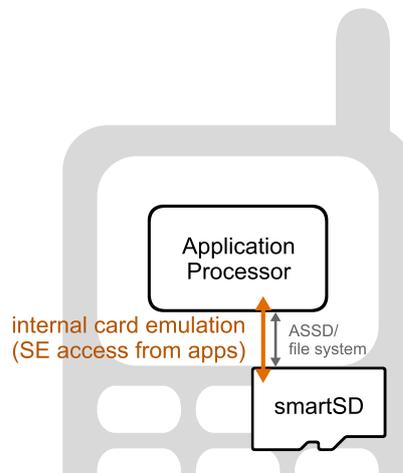

Figure 6: smartSD card in a mobile device without using NFC capabilities

As with the UICC scenario, a smartSD card may be used exclusively for providing the advantages of a secure element (secure storage, trusted execution environment) to apps running on the application processor in internal card emulation mode. In that case, the smartSD card is only connected to the application processor through the regular SD interface (see Fig. 6).

# 3. Open Mobile API

The Open Mobile API [18] is a specification created and maintained by SIMalliance, a non-profit trade association that aims at creating secure, open and interoperable mobile services. It defines a platform-independent middleware architecture between apps and secure elements on mobile devices, and specifies a programming language independent API for integrating secure element access into mobile applications.

## 3.1 Overall Architecture

The overall architecture of the Open Mobile API is shown in Fig. 7. The Open Mobile API consists of service APIs, a core transport API, and secure element provider driver modules.

The core component is the transport API which provides APDU-based connections to secure element applets. The transport API consists of four classes: `SEService`, `Reader`, `Session` and `Channel`. The `SEService` manages all secure element slots in a mobile device. Each secure element slot matches one secure element driver module and, therefore, corresponds to one secure element. Each slot is represented by an instance of the `Reader` class. The `Reader` class has methods to check the availability of the slot's secure element and to establish a session to the secure element. Once a session is established it is represented by a `Session` object. The `Session` class provides methods to obtain the answer-to-reset (ATR) of the secure element and to open APDU-based communication channels to applets on the secure element. Each communication channel is represented by a `Channel` object. The `Channel` class has a `transmit` method to exchange APDUs.

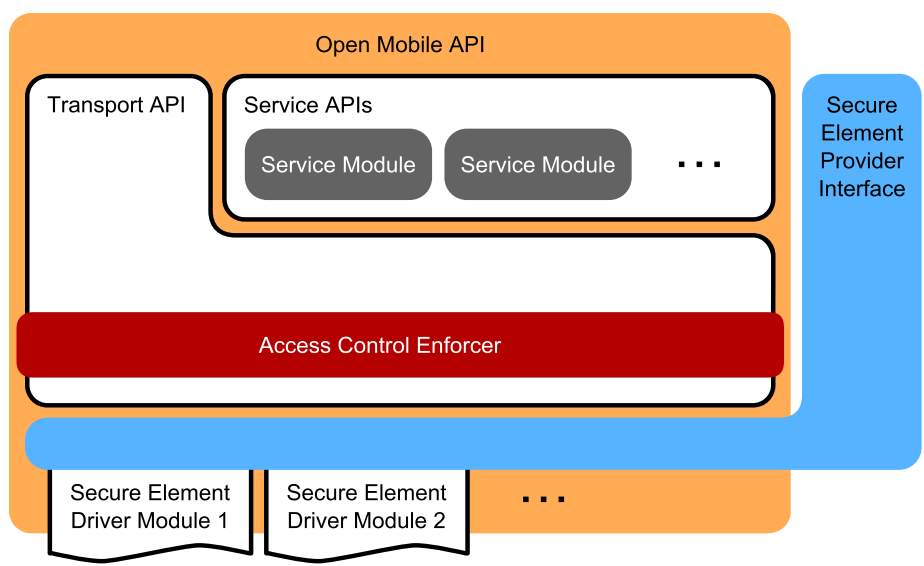

Figure 7: Architectural overview of the Open Mobile API [12, 18]

The service API is a collection of multiple service modules. Each module provides an application-specific abstraction of the transport layer. Thus, instead of low-level communication through APDUs, high-level methods can be defined for specific applications. For example, an authentication service API could provide methods for PIN code management and verification. Similarly, a file management service API could provide high-level methods to create, write, and read files in a smartcard file system.

An access control enforcer between the transport API and the secure element providers ensures that access restrictions to secure elements are obeyed. The security mechanism for access control enforcement is defined by GlobalPlatform's Secure Element Access Control specification [8].

The secure element provider interface defines an abstraction layer to add arbitrary secure element driver modules (each representing a secure element). These driver modules can be statically integrated into the system as well as dynamically loaded by third-party apps at runtime.

## 3.2 Secure Element Access Control

The access control enforcer is not part of the Open Mobile API itself. Instead, the Open Mobile API specification references to GlobalPlatform's Secure Element Access Control [8] specification. That specification defines a sophisticated security scheme for secure element APIs to prevent secure element access by unauthorized applications.

The architecture of the access control scheme is depicted in Fig. 8. The core component of the access control scheme is the access control enforcer. The enforcer resides within the secure element API (cf. Open Mobile API) and acts as a gatekeeper between apps on the mobile device and the secure element. Access control decisions are based on access control rules. Each rule defines access rights for a specific secure element applet based on its application identifier (AID) and a specific app or a specific group of apps on the mobile device based on their certificates. Moreover, rules may be applied to any secure element application without a specific AID and to mobile device apps without a specific certificate. Access rights can grant and deny access to all APDUs, to specific APDUs and to event notifications.

The access control enforcer reads the access control rules from a database on the secure element. Different methods for access to the database exist. The database can be a simple file, the access rule file (ARF), which is accessible through file access APDUs. The preferred way, however, is an access rule application (ARA). As the access rule databases may be distributed across multiple security domains owned and managed over-the-air by multiple entities, the ARA-M (ARA master) aggregates all these individually managed databases (ARA-C, ARA clients) and provides a standardized interface for the access control enforcer. Over-the-air (OTA) management of ARA-C databases is comparable to the OTA management of secure element applications.

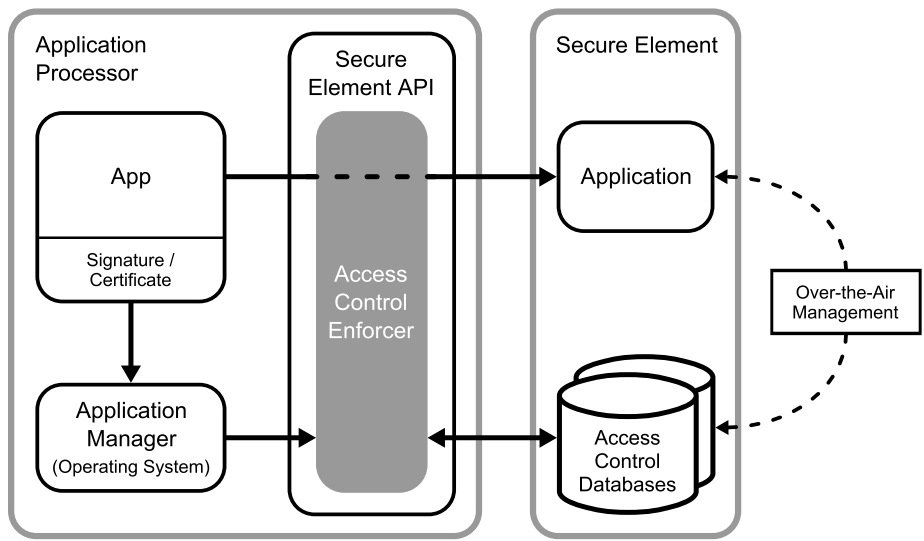

Figure 8: Secure Element Access Control Architecture [8]

When an app running on the application processor tries to access an applet on the secure element, the access control enforcer retrieves the app certificate from the application manager of the mobile device operating system and looks up the access rules for that certificate (or its certificate chain) and the selected applet AID. Based on these rules, access control is enforced for each transmitted APDU.

While the access policies themselves are stored on the secure element, the GlobalPlatform Secure Element Access Control delegates access control decisions to the operating system (or the secure element API) on the application processor of the mobile device. Hence, access control is delegated to a component with much less stringent security requirements. See [12] for a discussion of the implications of this delegation for secure element applications.

## 3.3 An Implementation: SEEK-for-Android Smartcard API

The project "Secure Element Evaluation Kit for the Android platform" (SEEK-for-Android, [4]) has been launched and is maintained by Giesecke & Devrient and provides the "Smartcard API" as an open-source implementation of the Open Mobile API specification for the Android operating system platform. The Smartcard API is released in the form of patches to the Android Open Source Platform (AOSP) as well as in the form of a series of GIT repositories[4] hosted on GitHub. Moreover, the SEEK-for-Android project provides add-ons for the Android SDK to integrate the Open Mobile API into Android applications and to simplify compilation against the Open Mobile API framework classes [5].

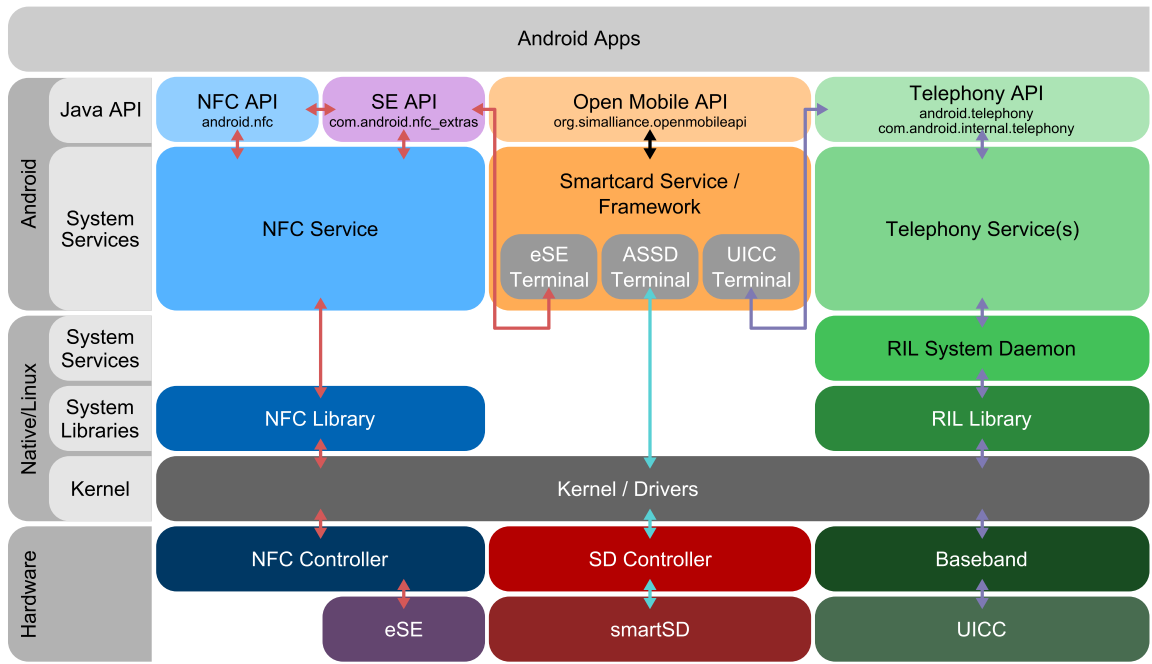

Figure 9: SEEK-for-Android Smartcard API (v3.1.0) within the Android platform

Figure 9 gives an overview of SEEK version 3.1.0 within the Android platform. The smartcard subsystem consists of the smartcard system service and the Open Mobile API framework. SEEK includes secure element interface modules ("terminals") for access to different forms of secure elements:

- An ASSD terminal provides access to a smartSD card under the assumption that the SD kernel driver supports ASSD commands.

- An eSE terminal provides access to an embedded secure element (eSE) that is accessible through Android's proprietary secure element API (`com.android.nfc_extras`).

- A UICC terminal provides access to the UICC through the telephony framework, given that the radio interface layer (RIL) supports standardized AT commands for APDU-based access to the UICC (cf. [3, 6, 7]).

---

[4]https://github.com/seek-for-android

## 3.4 Secure Element Provider Interface before Version 4.0.0

The Open Mobile API specification leaves the definition of the secure element provider interface open to the actual implementation of the Open Mobile API. Therefore, the secure element provider interface may vary between different implementations.

The SEEK-for-Android implementation supports two types of interface modules (so-called "terminals"):

1. terminals that are compiled into the smartcard system service, and

2. add-on terminals that can be added during runtime through Android application packages.

### 3.4.1 Integration as Compiled-In Terminal

The SEEK smartcard service can be bundled with SE terminal interface modules at compile-time. A terminal is a class that extends the abstract class `org.simalliance.openmobileapi.service.Terminal` and implements at least all the abstract interface methods. These terminals are typically located in the Java package namespace `org.simalliance.openmobileapi.service.terminals.*` (though this is no requirement).

### 3.4.2 Integration as Add-On Terminal

Terminal interface modules can also be added at runtime by installing Android application packages. SEEK automatically detects and adds such terminals. In order to be discoverable by SEEK, the application package name must start with "`org.simalliance.openmobileapi.service.terminals.`" The application package must contain at least one class that ends with the string "`Terminal`". A terminal class need not be located in any specific Java package namespace. However, all classes ending with the string "`Terminal`" must implement all terminal interface methods. As these interface methods are invoked through reflection, the terminal classes need not inherit from any specific Java interface or class as long as they implement the required interface methods.

### 3.4.3 Interface Methods

Terminal interface classes have to implement several methods that are used by the smartcard service to interact with the terminal module. These methods match the interface defined by the abstract base class `Terminal`.

- `public *Terminal(android.content.Context context)`:
  A constructor that takes an Android context as parameter.

- `public String getName()`:
  This method must return an identifying name for the terminal module, e.g. "UICC".

- `public boolean isCardPresent()`:
  This method must return `true` if the secure element of this terminal module is available and can be connected to.

- `public void internalConnect()`:
  This method is called before any connections to the secure element are established and can be used to initialize (e.g. power-up) the connection to the secure element.

- `public void internalDisconnect()`:
  This method is called when the secure element is no longer used (e.g. because all clients disconnected) and can be used for clean-up and to shutdown (e.g. power-down) the connection to the secure element.

- `public byte[] getAtr()`:
  This method must return the answer-to-reset of the secure element (or `null` if there is none or the ATR cannot be obtained for this type of SE). This method may be invoked before `internalConnect()` or after `internalDisconnect()`.

- `public int internalOpenLogicalChannel()`:
  This method is called to open a new logical channel. It must return the ISO/IEC 7816-4 logical channel number of the opened channel. If opening a logical channel without an explicit application identifier (AID) is not supported, this method must throw an `UnsupportedOperationException`. If there is no further logical channel available, a `MissingResourceException` must be thrown.

- `public int internalOpenLogicalChannel(byte[] aid)`:
  This method is called to open a new logical channel selecting a specific application by its AID. It must return the ISO/IEC 7816-4 logical channel number of the opened channel. If the application is not found, this method must throw a `NoSuchElementException`. If there is no further logical channel available, a `MissingResourceException` must be thrown.

- `public byte[] getSelectResponse()`:
  This method must return the response to the SELECT command that was used to open the basic channel or a logical channel (or `null` if a SELECT response cannot be retrieved).

- `public byte[] internalTransmit(byte[] command)`:
  This method is used to transmit a command APDU and to receive the corresponding response APDU on any channel using the ISO/IEC 7816-4 channel number as set in the CLA byte of the command APDU.

- `public void internalCloseLogicalChannel(int channel)`:
  This method is used to close a previously opened logical channel based on its ISO/IEC 7816-4 logical channel number.

## 3.5  Secure Element Provider Interface since Version 4.0.0

Starting with version 4.0.0, SEEK-for-Android uses a different concept to manage terminals. Terminals are no longer compiled into the smartcard system service. Instead, each terminal module is a separate Android service. The smartcard system service enumerates, manages and binds those services dynamically at run-time.

Typically, each terminal is encapsulated into a separate Android application package. This permits a minimization of the privileges required by the smartcard system service and by each terminal module. For instance, the smartcard system service no longer needs permissions to access the UICC, an eSE, or an ASSD, as this was the case with compiled-in terminals. Instead, there can be one terminal module application with only the permission to access the UICC, another one with only the permission to access the eSE, and another one with only the permission to access the ASSD.

### 3.5.1  Service Interface

A terminal module consists of an (exported) Android service component that filters for the intent `org.simalliance.openmobileapi.TERMINAL_DISCOVERY`. Moreover, binding to the service must require the system permission `org.simalliance.openmobileapi.BIND_TERMINAL`. This prevents other applications from bypassing the access control mechanisms of the smartcard system service by directly binding to the terminal module service components.

The following is an example of how such a terminal module service could be declared in an Android application manifest (`AndroidManifest.xml`):

```xml
<service android:name=".MyTerminal"
        android:label="MYTERMINAL"
        android:enabled="true"
        android:exported="true"
        android:permission=
          "org.simalliance.openmobileapi.BIND_TERMINAL">
    <intent-filter>
        <action android:name=
              "org.simalliance.openmobileapi.TERMINAL_DISCOVERY" />
    </intent-filter>
</service>
```

The terminal module service must implement the `ITerminalService` binder interface:

```
interface ITerminalService {
    OpenLogicalChannelResponse internalOpenLogicalChannel(
                    in byte[] aid,
                    in byte p2,
                    out SmartcardError error);
    void internalCloseLogicalChannel(int channelNumber,
                                     out SmartcardError error);
    byte[] internalTransmit(in byte[] command,
                            out SmartcardError error);
    byte[] getAtr();
    boolean isCardPresent();
    byte[] simIOExchange(in int fileID,
                         in String filePath,
                         in byte[] cmd,
                         out SmartcardError error);
    String getSeStateChangedAction();
}
```

This interface consists of the following methods:

- `isCardPresent`:
  This method must return `true` if the secure element of this terminal module
  is available and can be connected to.

- `getAtr`:
  This method must return the answer-to-reset of the secure element (or `null`
  if there is none or the ATR cannot be obtained for this type of SE).

- `internalOpenLogicalChannel`:
  This method is called to open a new logical channel selecting a specific applica-
  tion by its AID. It must return an `OpenLogicalChannelResponse` object that
  contains the ISO/IEC 7816-4 logical channel number of the opened channel
  and the response to the SELECT command that was used to open the logi-
  cal channel. Errors are reported by setting the reason in the `SmartcardError`
  object that is passed through the parameter `error`.

- `internalTransmit`:
  This method is used to transmit a command APDU and to receive the corre-
  sponding response APDU on any channel using the ISO/IEC 7816-4 channel
  number as set in the CLA byte of the command APDU. Errors are reported
  by setting the reason in the `SmartcardError` object that is passed through
  the parameter `error`.

- `internalCloseLogicalChannel`:
  This method is used to close a previously opened logical channel based on its
  ISO/IEC 7816-4 logical channel number. Errors are reported by setting the reason in the `SmartcardError` object that is passed through the parameter `error`.

- `simIOExchange`:
  This method is used to select and read files of the SIM/UICC file system and may be used by the access control enforcer. Errors are reported by setting the reason in the `SmartcardError` object that is passed through the parameter `error`.

- `getSeStateChangedAction`:
  This method must return the name of an Android broadcast intent action that is sent whenever the state of the secure element of this terminal module changes.

### 3.5.2 Differentiation between System and Add-on Terminals

While this version of the SEEK smartcard API no longer differentiates between compiled-in and add-on terminals, it differentiates between system terminals and other terminals. System terminals are terminals with a name of the form "SIM$x$", "eSE$x$", and "SD$x$" (where $x$ is a consecutively numbered index that starts at 1 for each terminal type). These terminals are listed first when obtaining a list of available terminals through the Open Mobile API. In order to prevent non-system applications from adding terminals to the top of the list, applications exporting system terminals are required to hold the system permission `org.simalliance.openmobileapi.SYSTEM_TERMINAL`. As a result, applications using the Open Mobile API can estimate if they talk to a system-provided or an add-on terminal.

## 3.6 Availability in Devices

As of today, many smartphones ship with an implementation of the Open Mobile API in their stock ROMs. Typically, these implementations give access to a UICC-based secure element. On some devices, they also provide access to an embedded secure element (e.g. on the Samsung Galaxy S3) or a smartSD card. When compared to SEEK, the vendor-specific implementations usually differ slightly in their behavior—even though all implementations that we discovered seem to be originally forked from SEEK (versions 3.1.0 or earlier).

For instance, we found differences in the access control mechanism. The SEEK implementation follows GlobalPlatform's Secure Element Access Control specification while some vendor-specific implementations only support (or prefer) an access rule file in a PKCS #15 file structure (cf. [8, 13]). Moreover, we found that some devices

Table 1: Open Mobile API support in existing devices

| Manufacturer | Model | Android version | API supported | Compiled-in terminals[a] | | | Add-on terminals |
|---|---|---|---|---|---|---|---|
| | | | | UICC | eSE | ASSD | |
| Fairphone | FP1 | 4.2.2 | no | — | — | — | — |
| HTC | One X | 4.2.2 | no | — | — | — | — |
| HTC | One mini 2 | 4.4.2 | yes | yes | n/a[b] | n/a[b] | yes |
| HTC | One (M8) | 5.0.2 | yes | yes | yes[c] | yes[c] | yes |
| Huawei | Ascend P7 | 4.4.2 | yes | yes | yes | yes | yes |
| Huawei | P8 lite | 4.4.2 | yes | yes | n/a[b] | n/a[b] | yes |
| LG | Nexus 4 | 5.1.1 | no | — | — | — | — |
| LG | Nexus 5 | 5.1.1 | no | — | — | — | — |
| LG | Optimus L5 II | 4.1.2 | yes | yes | n/a[b] | n/a[b] | n/a[b] |
| Motorola | RAZR i | 4.4.2 | yes | yes | yes | yes | yes |
| Motorola | Nexus 6 | 5.1.0 | yes | yes | no | no | yes |
| Motorola | Nexus 6 | 6.0.0 | no | — | — | — | — |
| OnePlus | One | 5.0.2 | no | — | — | — | — |
| Oppo | N5117 | 4.3 | yes | yes | yes[c] | no | yes |
| Samsung | Galaxy S3 | 4.1.2 | yes | yes | yes | no | yes[d] |
| Samsung | Galaxy S4 | 5.0.1 | yes | yes | yes | no | no |
| Samsung | Galaxy S4 mini | 4.4.2 | yes | yes | n/a[b] | n/a[b] | no |
| Samsung | Galaxy S5 | 4.4.2 | yes | yes | yes | no | no |
| Samsung | Galaxy S6 | 5.1.1 | yes | yes | yes | no | no |
| Samsung | Xcover 3 | 4.4.4 | yes | yes | no[e] | no | no |
| Sony | Xperia M2 Aqua | 4.4.4 | yes | yes | n/a[b] | n/a[b] | n/a[b] |
| Sony | Xperia Z3 Compact | 5.0.2 | yes | yes | yes | yes | no |

[a]Implementations of these terminals are provided as part of the smartcard service application package. Access to these terminals through the Open Mobile API has not been tested.

[b]This aspect has not been analyzed/tested.

[c]These terminals are provided through a separate instance of the smartcard service with the package name `com.nxp.nfceeapi.service`. Access to these terminals has not been tested.

[d]In addition to the add-on terminal interface defined by SEEK, add-on terminals must also implement the methods `public boolean isChannelCanBeEstablished()` (used to check if further logical channels to the secure element can be opened) and `public void setCallingPackageInfo (String packageName, int uid, int pid)` (used to inform the add-on terminal about the calling application (package name, user ID, and process ID) before opening channels and exchanging APDUs).

[e]An empty stub implementation is provided as part of the smartcard service application package.

do not support add-on terminals[5].

Table 1 gives an overview of analyzed devices and their support for the Open Mobile API. Due to the standardized API definition of the Open Mobile API, apps compiled against the SEEK smartcard API framework seamlessly integrate with these vendor-specific implementations shipped in stock ROMs.

---

[5]Possibly due to the security implications of add-on terminals in SEEK versions 3.1.0 and earlier (cf. CVE-2015-6606).

# 4. Reverse-Engineering Android Applications

We assembled a toolbox consisting of a number of existing applications for analyzing, decompiling, and manipulating Android application packages and libraries. We used this toolbox to evaluate the differences between the implementations of the Open Mobile API on various Android devices, and, specifically, to reverse-engineer and analyze how the Open Mobile API implementation on the Samsung Galaxy S3 accesses the UICC.

## 4.1 Tools

Our toolbox consists of the following freely available applications:

- The `adb` command (Android Debug Bridge) from the Android SDK platform tools is used to pull the Android framework, application packages and libraries from the system partition of existing Android devices.

- *Apktool*[6] is used to assemble the framework files from the pulled files and to extract resources (including the `AndroidManifest.xml` file) from (optimized) Android application packages.

- A combination of *smali* and *baksmali*[7] is used to transform optimized Dalvik executables (`.odex` files) into non-optimized Dalvik executables (`.dex` files). In other words, `.odex` files containing the optimized executable program code stripped off Android application packages when shipping them on the system partition of an Android device are translated into Dalvik bytecode packed into a `classes.dex` file as it is usually embedded into stand-alone Android application packages. Moreover, baksmali is used to disassemble Dalvik bytecode.

- The tool *oat2dex*[8] is used to extract (and possibly de-optimize) Dalvik executables from ahead-of-time compiled executables (`.oat` files) for the Android runtime (ART).

- The tool *dex2jar*[9] is used to translate Dalvik bytecode from Dalvik executables (`.dex` files) and Android application packages (`.apk` files) into Java bytecode.

- The Java decompiler *JD-GUI*[10] is used to decompile Java bytecode into Java source code.

---

[6] http://ibotpeaches.github.io/Apktool/
[7] https://github.com/JesusFreke/smali
[8] https://github.com/testwhat/SmaliEx
[9] https://github.com/pxb1988/dex2jar
[10] http://jd.benow.ca/

## 4.2 Using the Tools

In this section, we demonstrate how we used these tools to extract, decompile, and analyze the Android system files (framework and system applications) of a Samsung Galaxy S3 running its stock firmware based on Android 4.1.2.

### 4.2.1 Downloading Files from the Device

As a first step, we pulled the framework files and the Android application packages from the system partition of the Samsung Galaxy S3 using the Android Debug Bridge tool (`adb`). In order to allow access through the `adb` tool, we activated the developer options and enabled the Android Debug Bridge interface (*Settings → Developer options → Android debugging*) on the device.

The following commands were used to pull all the framework files:

```
$ adb pull /system/framework ./gs3/framework
$ adb pull /system/app/minimode-res.apk ./gs3/framework
```

Based on our experience with other devices, the actual location of the framework files (specifically of files outside `/system/framework` containing resources required to de-optimize DEX files) seems to vary.

For our analysis, the most interesting framework files were:

- `framework-res.apk`, `twframework-res.apk`, and `minimode-res.apk`: These files contain the resources necessary to de-optimize `.odex` files.

- `framework.odex`: This file contains the implementation of the Android API framework.

- `framework2.odex`: This file contains extensions to the Android API framework.

- `org.simalliance.openmobileapi.odex`: This file contains the implementation of the Open Mobile API framework.

- `com.android.nfc_extras.odex`: This file contains the implementation of the API framework for access to an embedded secure element.

The following command was used to pull all the Android applications integrated into the system partition:

```
$ adb pull /system/app ./gs3/app
```

As with the framework files, the actual location may vary between different devices. Specifically, starting with Android 4.4, an additional directory `/system/priv-app` contains privileged system applications that can obtain "signatureOrSystem" permissions without being signed with the platform key. Moreover, the structure of these directories has significantly changed starting with Android 5.0.

For our analysis, the most interesting Android application files were:

- `SmartcardService.apk` and `SmartcardService.odex`: These files contain the smartcard system service implementation.

- `SecPhone.apk` and `SecPhone.odex`: These files contain the telephony system service implementation.

- `Nfc.apk` and `Nfc.odex`: These files contain the NFC system service implementation.

### 4.2.2  Preparing the Framework Files

Certain framework resource files are necessary to de-optimize and decode Android application packages and optimized framework files. Their names typically end in "`-res.apk`". Apktool is used to collect and prepare those files:

```
$ java -jar ./bin/apktool.jar if -p ./fw \
    ./gs3/framework/framework-res.apk
$ java -jar ./bin/apktool.jar if -p ./fw \
    ./gs3/framework/minimode-res.apk
$ java -jar ./bin/apktool.jar if -p ./fw \
    ./gs3/framework/twframework-res.apk
```

### 4.2.3  De-optimizing Dalvik Executables

After collecting the framework resources, we can de-optimize optimized ("odex-ed") Dalvik executables of both, framework files and apps. The tool *baksmali* is used to de-optimize and disassemble the executable code into source files based on the *smali* assembly language syntax:

```
$ java -jar ./bin/baksmali.jar -a16 -d ./fw \
    -o ./gs3/decoded/SmartcardService.deodexed \
    -x ./gs3/app/SmartcardService.odex
```

The above command outputs the source files into the directory `./gs3/decoded/SmartcardService.deodexed`. We can then use the tool *smali* to assemble those source files into Dalvik bytecode (Dalvik executable). The following command generates the Dalvik executable `SmartcardService.dex`:

```
$ java -jar ./bin/smali.jar -a16 \
```

```
        -o ./gs3/decoded/SmartcardService.dex \
    ./gs3/decoded/SmartcardService.deodexed
```

### 4.2.4 Unpacking Application Packages

Android application packages can be extracted using *Apktool*. For optimized APKs, the collected framework resources are used for de-optimization. We use the following command:

```
$ java -jar ./bin/apktool.jar d -p ./fw \
    -o ./gs3/decoded/SmartcardService.source \
    ./gs3/app/SmartcardService.apk
```

The above command outputs all resource files (`AndroidManifest.xml`, string resources, etc.) in their source form (and for non-optimized application packages also the Dalvik executable `classes.dex`) into the directory `./gs3/decoded/Smartcard-Service.source`.

### 4.2.5 Converting Dalvik Bytecode to Java Bytecode

As an intermediate step to decompiling Dalvik executables into Java source code, we translate Dalvik bytecode into Java bytecode using the tool *dex2jar*. This tool can either be used on a Dalvik executable (`.dex` file) or directly on an Android application package (`.apk` file):

```
$ d2j-dex2jar -o ./gs3/decoded/SmartcardService.jar \
    ./gs3/decoded/SmartcardService.dex
```

### 4.2.6 Decompiling Java Bytecode

Finally, the Java bytecode can be decompiled using a standard Java decompiler. We used JD-GUI to decompile the bytecode into its source form:

```
$ java -jar ./bin/jd-gui.jar \
    ./gs3/decoded/SmartcardService.jar
```

Figure 10 shows exemplary results of the decompiler.

## 4.3 Interpreting Decompiled Code: Results

Our reverse-engineering toolchain has some limitations when it comes to generating Java source code. Many parts of the Dalvik program code can be translated into working (or at least easily readable) Java source code. However, sometimes, specifically when it comes to certain code constructs, translation from Dalvik bytecode to Java source code produces code that is difficult to read or even completely erroneous.

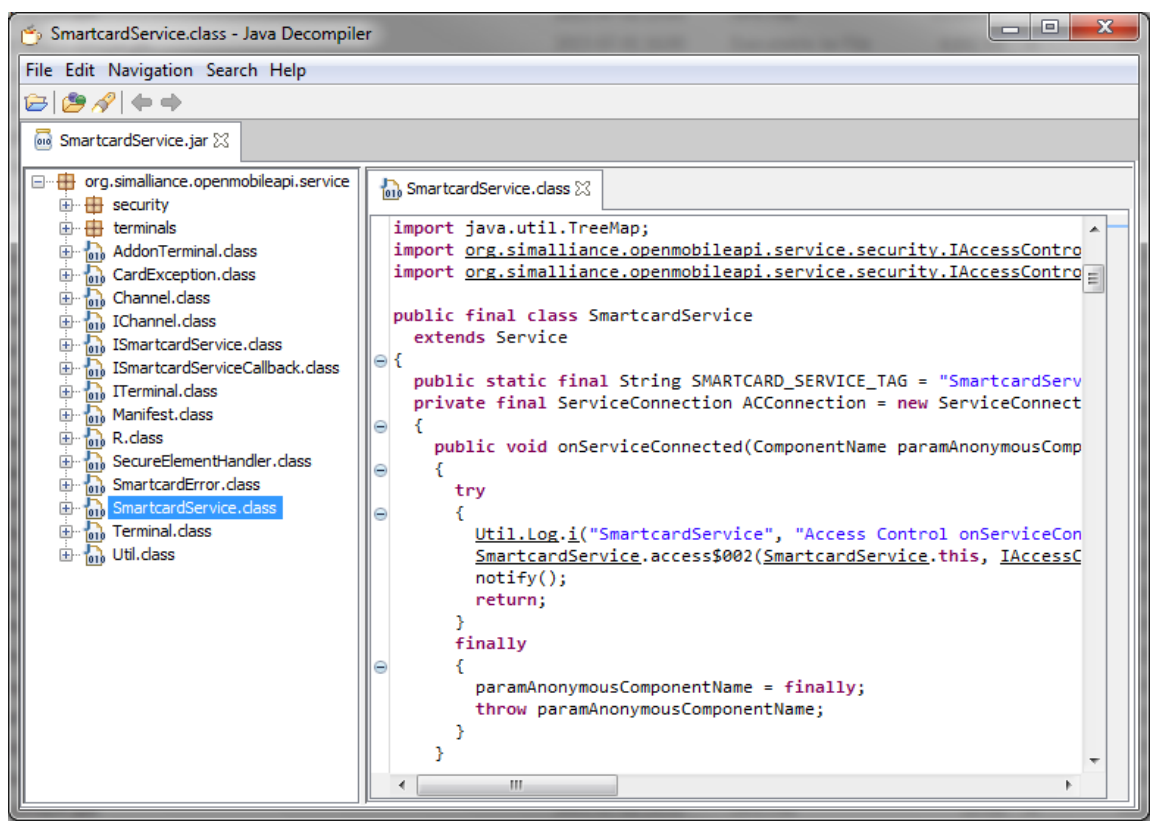

Figure 10: JD-GUI

For example, the following method, that converts a byte array into a string of hexadecimal digits, is translated into working Java source code:

```
1   public static String bytesToHexString(byte[] paramArrayOfByte) {
2       if (paramArrayOfByte == null) {
3           return null;
4       }
5       StringBuilder localStringBuilder =
6               new StringBuilder(paramArrayOfByte.length * 2);
7       int i = 0;
8       while (i < paramArrayOfByte.length) {
9           localStringBuilder.append("0123456789abcdef"
10                  .charAt(paramArrayOfByte[i] >> 4 & 0xF));
11          localStringBuilder.append("0123456789abcdef"
12                  .charAt(paramArrayOfByte[i] & 0xF));
13          i += 1;
14      }
```

```
15        return localStringBuilder.toString();
16 }
```

Only the names of parameters and local variables, still present in the disassembled Dalvik bytecode (see appendix A.1), get lost during the translation into Java code. Consequently, conditional statements and simple loops do not pose a problem to the decompilation toolchain.

In some cases, however, the Java decompiler produces readable code with minor issues that would prevent the generated source code from compiling (cf. appendix A.2 for the disassembled Dalvik bytecode):

```
1  public int openIccLogicalChannel(String paramString) {
2      if (DBG_ENG) {
3          Log.d("PhoneInterfaceManager",
4                  ">␣openIccLogicalChannel␣" + paramString);
5      }
6      paramString = (Integer)sendRequest(14,
7                              new IccOpenChannel(paramString));
8      if (DBG_ENG) {
9          Log.d("PhoneInterfaceManager",
10                 "<␣openIccLogicalChannel␣" + paramString);
11     }
12     return paramString.intValue();
13 }
```

In the above code, `paramString`, defined as `String`, is treated as an `Integer` object starting on line 6.

Unfortunately, certain code constructs in Dalvik executables result in completely unreadable and even wrong Java source code. For instance, this seems to be the case with `switch` statements like in the following example:

```
1  public IccException getException() {
2      if (success()) {
3          return null;
4      }
5      switch (this.sw1) {
6          default:
7              return new IccException("sw1:" + this.sw1 +
8                                      "␣sw2:" + this.sw2);
9      }
10     if (this.sw2 == 8) {
11         return new IccFileTypeMismatch();
12     }
13     return new IccFileNotFound();
14 }
```

The generated Java code only has a `default` label for the `switch` statement. Hence,

`new IccException(...)` (see line 8) would be returned regardless of the value of `sw1`. Lines 10–13 are unreachable.

However, a glance at the disassembled Dalvik bytecode (see appendix A.3) reveals that a packed-switch statement was incorrectly transformed. The packed-switch statement looks like this:

```
1  iget v0, p0, Lcom/android/internal/telephony/IccIoResult;->sw1:I
2  packed-switch v0, :pswitch_data_48
3
4  new-instance v0, Lcom/android/internal/telephony/IccException;
   (...)
6  goto :goto_7
7
8  :pswitch_35
   (...)
10 goto :goto_7
   (...)
12 :pswitch_data_48
13 .packed-switch 0x94
14     :pswitch_35
15 .end packed-switch
```

Therefore, only the default branch (lines 4–6 in the above listing) of the packed-switch statement was properly translated into Java code. The **case 0x94** statement (see the switching table on lines 13–15) was not translated into Java code. Consequently, the code starting at the label `:pswitch_35` (which maps to lines 10–13 of the generated Java source code) is seemingly unreachable.

Hence, the correct Java source code corresponding to the Dalvik bytecode would be:

```java
1  public IccException getException() {
2      if (success()) {
3          return null;
4      }
5      switch (this.sw1) {
6          case 0x94:
7              if (this.sw2 == 8) {
8                  return new IccFileTypeMismatch();
9              }
10             return new IccFileNotFound();
11
12         default:
13             return new IccException("sw1:" + this.sw1 +
14                                     " sw2:" + this.sw2);
15     }
16 }
```

# 5. SEEK on the Galaxy S3

Using our reverse-engineering toolbox, we disassembled and decompiled the Open Mobile API framework, the smartcard service, and all components that the smartcard service uses to access secure elements on our Samsung Galaxy S3.

## 5.1 Open Mobile API Framework

The Open Mobile API framework is located in `/system/framework/org.simalliance.openmobileapi.odex`. It follows the Open Mobile API specification and connects to the smartcard system service in order to access secure elements.

## 5.2 Smartcard System Service

The smartcard system service (`org.simalliance.openmobileapi.service`) is located in the application package `/system/app/SmartcardService.*`. This application contains two interesting classes for access to secure elements:

1. `org.simalliance.openmobileapi.service.terminals.SmartMxTerminal`, for access to an embedded secure element through a Samsung-specific variation of the Android API for embedded secure elements, and

2. `org.simalliance.openmobileapi.service.terminals.UiccTerminal`, for access to a UICC-based secure element through the telephony system service.

## 5.3 UICC Terminal Interface

As we were specifically interested in accessing the UICC, we further analyzed the UICC terminal interface class. When this class is instantiated, it obtains a handle (Android binder interface `com.android.internal.telephony.ITelephony`) to access the telephony service:

```
manager = ITelephony.Stub.asInterface(
                  ServiceManager.getService("phone"));
```

The implementation of the `UiccTerminal` class then accesses various methods of the `ITelephony` interface:

- The method `getAtr()` is used to obtain the ATR of the UICC-based secure element:

```
byte[] atr = manager.getAtr();
```

- The method `openIccLogicalChannel()` is used to open a new logical channel selecting a specific smartcard application (based on its AID, encoded as a hexadecimal string):

```
int channelId = manager.openIccLogicalChannel(aidStr);
```

- The method `getSelectResponse()` is used to obtain the response to the SE-LECT APDU command used to select the smartcard application on a previously opened logical channel:

```
byte[] selectResponse = manager.getSelectResponse();
```

- The method `closeIccLogicalChannel()` is used to close a previously opened logical channel:

```
boolean success = manager.closeIccLogicalChannel(
                         channelId[channel]);
```

- The method `getLastError()` is used to obtain an error code in case opening a logical channel failed:

```
int result = manager.getLastError();
```

- The method `transmitIccLogicalChannel()` is used to exchange APDUs over any logical channel (including the basic channel):

```
String response = manager.transmitIccLogicalChannel(
                         cla & 0x0FC,
                         ins,
                         channelId[cla & 0x003],
                         p1, p2,
                         len, dataStr));
```

The logical channel information is removed from the class byte (CLA) of the APDU. Instead, the channel identifier obtained by `openIccLogicalChannel()` is passed to this method. A channel identifier of 0 is used for the basic channel. Due to the way how channel identifiers are mapped to channel numbers, only up to four logical channels are supported (cf. [10], which would allow up to 20 logical channels).

The values of `len` and `data` depend on the type of the APDU (cf. [10]):

1. Case 1 (only command header): `len` is set to $-1$ and `dataStr` is set to `null`.

2. Case 2 (response data, no command data): `len` is set to the value of the Le field and `dataStr` is set to `null`. Only a single-byte Le field is properly mapped to `len`. In case of a multi-byte Le field, the remaining bytes would be treated as `dataStr`.

3. Case 3 (command data, no response data): `len` is set to the value of the Lc field and `dataStr` contains the command DATA field (encoded as hexadecimal string). Only a single-byte Lc field is properly mapped to `len`. In case of a multi-byte Lc field, the remaining bytes would be treated as part of `dataStr`.

4. Case 4 (command data, response data): `len` is set to the value of the Lc field and `dataStr` contains the command DATA field and the Le field (encoded as hexadecimal string). As with case 3, all but the first byte of a multi-byte Lc field would be treated as part of `dataStr`.

Besides those methods, the system property "gsm.sim.state" is used to determine if a UICC is inserted into the device and ready to be accessed:

```java
public boolean isCardPresent() throws CardException {
    return "READY".equals(SystemProperties.get(
                          "gsm.sim.state"));
}
```

## 5.4 Telephony System Service

The Binder interface `ITelephony` is used to establish an IPC connection to the telephony system service. This system service (`com.android.phone`) is encapsulated in the application package `/system/app/SecPhone.*`.

The `ITelephony` interface is implemented by the class `com.android.phone.Phone-InterfaceManager` (see appendix B.1 for the implementation and Fig. 11 for the class diagram). All interface methods used by the UICC terminal module invoke the

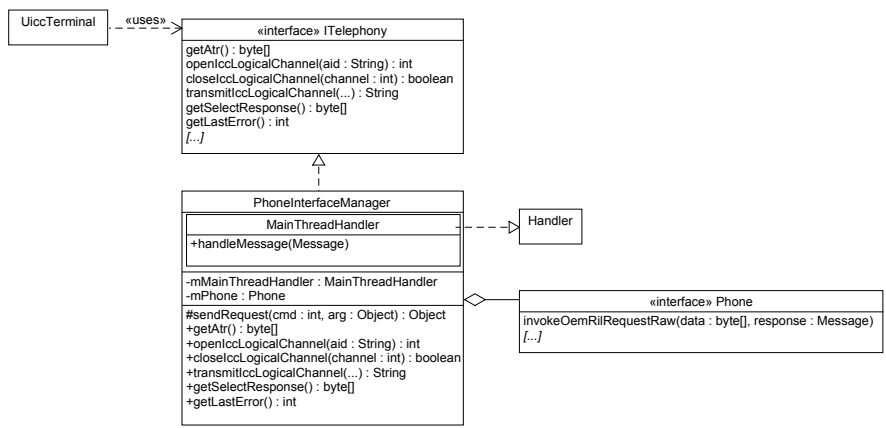

Figure 11: Class diagram for the relevant classes of SEEK and the telephony system service

`sendRequest()` method passing a command code and a command parameter. This method, in turn, posts a request (based on the command and its parameters) into a message queue and waits for the request to be processed. The message queue is processed by an instance of the class `MainThreadHandler` implemented inside the `PhoneInterfaceManager`. The message handler converts each request into a byte array that is passed to the radio interface (RIL system daemon) using the method `invokeOemRilRequestRaw()` (RIL command `RIL_REQUEST_OEM_HOOK_RAW` [= 59]). Similarly, the message handler processes raw responses posted by the radio interface into response values expected by the high-level interface methods.

### 5.4.1 RIL_REQUEST_OEM_HOOK_RAW

All commands related to access to the UICC are exchanged by means of the RIL command `RIL_REQUEST_OEM_HOOK_RAW`. Commands seem to follow a common frame structure:

| Command (2 bytes) | | Length (2 bytes) | Parameters (optional, $n$ bytes) |
|---|---|---|---|
| `0x15` | `0x0X` | $4 + n$ | $\cdots$ |

Where *command* is a 2-byte field of the form `0x150X` denoting the proprietary ("raw") radio interface command, `length` is a 2-byte field denoting the overall length of the command frame (including the *command* and *length* fields), and *parameters* contains zero or more command arguments. Multi-byte integer values are transmitted in big-endian format.

Responses are either encoded as byte arrays with command-specific formats or as RIL error codes. The following error codes have values deviating from those defined by SEEK-for-Android and AOSP:

- `INVALID_PARAMETER` uses the value 27,

- `MISSING_RESOURCE` uses the value 29, and

- `NO_SUCH_ELEMENT` uses the value 30.

When we later reimplemented UICC access for the Galaxy S3 on top of CyanogenMod, we found that the above error code mapping (obtained by reverse-engineering Samsung's implementation of the framework class `com.android.internal.telephony.RILConstants`) is wrong. Instead, the error code 29 is returned in cases where SEEK expects a `NO_SUCH_ELEMENT` error and the error code 30 is returned in cases where SEEK expects a `MISSING_RESOURCE` error. Consequently, the correct numbering of the RIL error codes is

- `INVALID_PARAMETER` = 27,

- `MISSING_RESOURCE` = 30, and

- `NO_SUCH_ELEMENT` = 29.

### 5.4.2 Getting the Answer-to-Reset

The command code used to obtain the answer-to-reset (ATR) of the UICC is `0x150D`. The implementation is listed in appendix B.1 on lines 418–461. The following command frame is passed to `invokeOemRilRequestRaw()`:

| Command (2 bytes) | | Length (2 bytes) | |
|------|------|------|------|
| 0x15 | 0x0D | 0x00 | 0x04 |

Upon success, this command returns the length of the ATR (1 byte), followed by one byte with unknown function that is ignored by the Samsung implementation, followed by the bytes of the ATR:

| ATR Len. (1 byte) | Unknown (1 byte) | ATR ($n$ bytes) |
|------|------|------|
| $n$ | 0xXY | $\cdots$ |

### 5.4.3 Opening a Logical Channel

The command code used to open a new logical channel is `0x1509`. The implementation is listed in appendix B.1 on lines 265–355. The command takes the application identifier (AID) of a smartcard application—that should be selected on the new logical channel—as parameter. The following command frame is passed to `invokeOemRilRequestRaw()`:

| Command (2 bytes) | | Length (2 bytes) | AID (n bytes) |
|------|------|------|------|
| 0x15 | 0x09 | $4 + n$ | $\cdots$ |

Upon success, this command returns the length of the logical channel ID (1 byte), followed by the logical channel ID, followed by the length of the response to the SELECT (by AID) command, followed by the actual response to the SELECT (by AID) command:

| ID Len. (1 byte) | Channel ID ($n$ bytes) | Resp. Len. (1 byte) | SELECT Response ($m$ bytes) |
|------|------|------|------|
| $n$ | $\cdots$ | $m$ | $\cdots$ |

In case of an error, one of the error codes `MISSING_RESOURCE` (indicating that all available logical channels are in use) or `NO_SUCH_ELEMENT` (indicating that the application AID could not be selected) is returned.

### 5.4.4 Closing a Logical Channel

The command code used to close a previously opened logical channel is `0x150A`. The implementation is listed in appendix B.1 on lines 357–416. The command takes the channel ID (cf. response to opening a logical channel) as parameter. The following command frame is passed to `invokeOemRilRequestRaw()`:

| Command (2 bytes) | | Length (2 bytes) | | Channel ID (4 bytes) |
|------|------|------|------|------|
| 0x15 | 0x0A | 0x00 | 0x08 | · · · |

The response does not contain any data. In case channel ID does not reference a previously opened logical channel, the error code `INVALID_PARAMETER` is returned.

### 5.4.5 Exchanging an APDU Command on the Basic Channel

The command code used to exchange an APDU command on the basic channel (i.e. logical channel 0) is `0x1508`. The implementation is listed in appendix B.1 on lines 158–263. The command takes the header and the optional body field of the command APDU as parameters. The body field consists of a length byte (P3) and the data field. P3 matches the `len` parameter of the `transmitIccLogicalChannel()` method and, therefore, maps to the Le byte (cf. [10]) if there is no command data field or the Lc byte if the command data field is not empty. We did not test if it is possible to wrap extended length APDUs or case 4 APDUs (cf. APDU cases in section 5.3) with this command by putting the remaining bytes of the length fields in the data part. This is what would happen if such an APDU is passed in though the SEEK smartcard service. It could also be the case that a mapping procedure similar to that defined in [9] for the T=0 transmission protocol needs to be implemented. This is currently not the case with SEEK.

The following command frame is passed to `invokeOemRilRequestRaw()`:

| Command (2 bytes) | | Length (2 bytes) | APDU Header (4 bytes) | | | | APDU Data (optional, 1 + n bytes) | |
|------|------|------|------|------|------|------|------|------|
| 0x15 | 0x08 | 8 or 9 + n | CLA | INS | P1 | P2 | P3 | · · · |

Upon success, this command returns the corresponding response APDU (cf. [10]):

| Response Data (n bytes) | Status (2 bytes) | |
|------|------|------|
| · · · | SW1 | SW2 |

In case a malformed command APDU is passed as parameter, the error code `INVALID_PARAMETER` is returned.

### 5.4.6 Exchanging an APDU Command on a Logical Channel

Two different command codes are used to exchange an APDU command on a logical channel (other than channel 0): `0x150B` and `0x150C`. The implementation is listed in appendix B.1 on lines 158–263.

The first version of this command (command code `0x150B`) is used when there is at least a length field (*Lc* or *Le*) included into the command APDU. In this case, the command takes the APDU header, the length byte (P3, cf. section 5.4.5), the channel ID (cf. response to opening a logical channel), and the optional APDU data field as parameters. Again, we did not test if it is possible to wrap extended length APDUs or case 4 APDUs with this command. The following command frame is passed to `invokeOemRilRequestRaw()`:

| Command (2 bytes) | | Length (2 bytes) | APDU Header (4 bytes) | | | | P3 (1 B) | Channel ID (4 bytes) | APDU Data (optional, $n$ bytes) |
|---|---|---|---|---|---|---|---|---|---|
| 0x15 | 0x0B | $13 + n$ | CLA | INS | P1 | P2 | P3 | · · · | · · · |

The second version of this command (command code `0x150C`) is used for case 1 APDUs (i.e. APDUs that have no command data field and expect no response data, cf. [10]). In this case, the command takes the APDU header and the channel ID (cf. response to opening a logical channel) as parameters. The following command frame is passed to `invokeOemRilRequestRaw()`:

| Command (2 bytes) | | Length (2 bytes) | | APDU Header (4 bytes) | | | | Channel ID (4 bytes) |
|---|---|---|---|---|---|---|---|---|
| 0x15 | 0x0C | 0x00 | 0x0C | CLA | INS | P1 | P2 | · · · |

Upon success, these commands return the corresponding response APDU (cf. [10]):

| Response Data ($n$ bytes) | Status (2 bytes) | |
|---|---|---|
| · · · | SW1 | SW2 |

In case a malformed command APDU or an invalid channel ID are passed as parameters, the error code `INVALID_PARAMETER` is returned.

# 6. Adding UICC Terminal Support to CyanogenMod

Our analysis of the Samsung Galaxy S3 revealed that we need to modify the RIL implementation of the telephony system service in order to permit access to the UICC by sending proprietary command blobs to the RIL system daemon.

## 6.1 CyanogenMod 11.0 for the Samsung Galaxy S3

As a first step towards integrating SEEK into CyanogenMod, the CyanogenMod sources for the Samsung Galaxy S3 (i9300) need to be obtained. We chose version 11.0 as this is the most recent version available for that device.

First, a new directory for checking out CyanogenMod has to be created:

```
$ mkdir cm-11.0
$ cd cm-11.0
$ repo init \
    -u https://github.com/CyanogenMod/android.git \
    -b stable/cm-11.0
```

To obtain the configuration of the source tree that was used to generate a specific CyanogenMod build (e.g. cm-11-20151004-NIGHTLY-i9300), the build manifest from the respective over-the-air update `.zip` file needs to be extracted and used to initialize the local repository:

```
$ unzip -p ../cm-rel/cm-11-20151004-NIGHTLY-i9300.zip \
    system/etc/build-manifest.xml >\
    .repo/manifests/cm-11-20151004-NIGHTLY-i9300.xml
$ repo init -m cm-11-20151004-NIGHTLY-i9300.xml
```

Finally, the sources and pre-built files can be downloaded:

```
$ repo sync
$ cd vendor/cm
$ ./get-prebuilts
$ cd ../..
```

For building the complete Android system, some additional proprietary files need to be extracted from an existing Galaxy S3 with that CyanogenMod version. Alternatively, these files can also be extracted from the respective over-the-air update `.zip` file (e.g. `cm-11-20151004-NIGHTLY-i9300.zip`).

## 6.2 Patches to Include SEEK-for-Android

In CyanogenMod version 10.2 and earlier, patches provided by the SEEK project had to be applied to the CyanogenMod source tree. Since version 11.0, SEEK-for-Android is integrated into the CyanogenMod source tree. Therefore, these patches no longer need to be applied. However, the SEEK smartcard system service and the API framework must be included into the build process for the Galaxy S3 (i9300). This can be accomplished by adding the following lines to the file `device/samsung/i9300/i9300.mk`:

```
PRODUCT_PACKAGES += \
    org.simalliance.openmobileapi \
    org.simalliance.openmobileapi.xml \
    SmartcardService

PRODUCT_PROPERTY_OVERRIDES += \
    persist.nfc.smartcard.config=SIM1
```

Moreover, the smartcard service project is excluded from the build-process by default. It can be included by adding the following line to the file `device/samsung/i9300/BoardConfig.mk`:

```
TARGET_ENABLE_SMARTCARD_SERVICE := true
```

Further, the version of the smartcard system service included into CyanogenMod is based on a non-standard implementation of the Android eSE API. This non-standard eSE API is not available on the Galaxy S3. Consequently, the references to that API need to be removed.

First, in the file `packages/apps/SmartCardService/Android.mk`, the entry `com.android.qcom.nfc_extras` must be removed from the variable `LOCAL_JAVA_LIBRARIES`:

```
LOCAL_JAVA_LIBRARIES := core framework \
    org.simalliance.openmobileapi
```

Second, the eSE terminal implementation has to be removed by deleting the file `SmartMxTerminal.java` (in `packages/apps/SmartCardService/src/org/simalliance/openmobileapi/service/terminals/`). Finally, all references to the class `com.android.qcom.nfc_extras.NfcQcomAdapter` need to be removed from the file `SmartcardService.java` (in `packages/apps/SmartCardService/src/org/simalliance/openmobileapi/service/`).

Moreover, the ASSD terminal in not necessary for our purpose and can also be removed by deleting the file `ASSDTerminal.java` (in `packages/apps/SmartCardService/src/org/simalliance/openmobileapi/service/terminals/`).

## 6.3 Enabling UICC Access through SEEK

This section describes the changes necessary to enable access to the UICC through SEEK in CyanogenMod 11.0 on a Samsung Galaxy S3 (see Fig. 12 for an overview of involved classes, methods, and fields). A complete patch set with all these changes to the CyanogenMod 11.0 source tree is available on our website[11].

### 6.3.1 Radio Interface Layer

As the changes necessary to enable access to the UICC seem to be specific to the Exynos4 system-on-chip that is used in the Samsung Galaxy S3, the best place to add these platform-specific additions would be the Exynos4 RIL implementation located in `frameworks/opt/telephony/src/java/com/android/internal/telephony/SamsungExynos4RIL.java`. With this approach, the changes only apply to real hardware that uses the Exynos4 RIL implementation while the same source tree can be compiled for other target devices or the emulator using the existing SEEK emulator extensions without being influenced by our modifications.

SEEK already adds a standard implementation for UICC access (usable with the emulator) to the default RIL implementation. This standard implementation consists of the following four methods that invoke RIL-specific commands:

- `iccExchangeAPDU(int cla, int command, int channel, int p1, int p2, int p3, String data, Message result)` for exchanging APDU commands with the UICC,

- `iccOpenChannel(String aid, Message result)` for opening a logical channel to an applet on the UICC,

- `iccCloseChannel(int channel, Message result)` for closing a previously opened logical channel, and

- `iccGetAtr(Message result)` for retrieving the ATR of the UICC.

Consequently, the standard implementation of these methods has to be overridden in the device-specific subclass `SamsungExynos4RIL`. As all of these commands are encapsulated in the RIL command `RIL_REQUEST_OEM_HOOK_RAW`, our implementation defines constants to tag each of these individual commands. This permits mapping responses to `RIL_REQUEST_OEM_HOOK_RAW` to a specific command while processing the response message handler:

```
private static final int RIL_OEM_HOOK_RAW_EXCHANGE_APDU = 1;
private static final int RIL_OEM_HOOK_RAW_OPEN_CHANNEL  = 2;
private static final int RIL_OEM_HOOK_RAW_CLOSE_CHANNEL = 3;
private static final int RIL_OEM_HOOK_RAW_GET_ATR       = 4;
```

---

[11] https://usmile.at/sites/default/files/blog/uicc_on_seek_on_cm11_0_gs3.patch

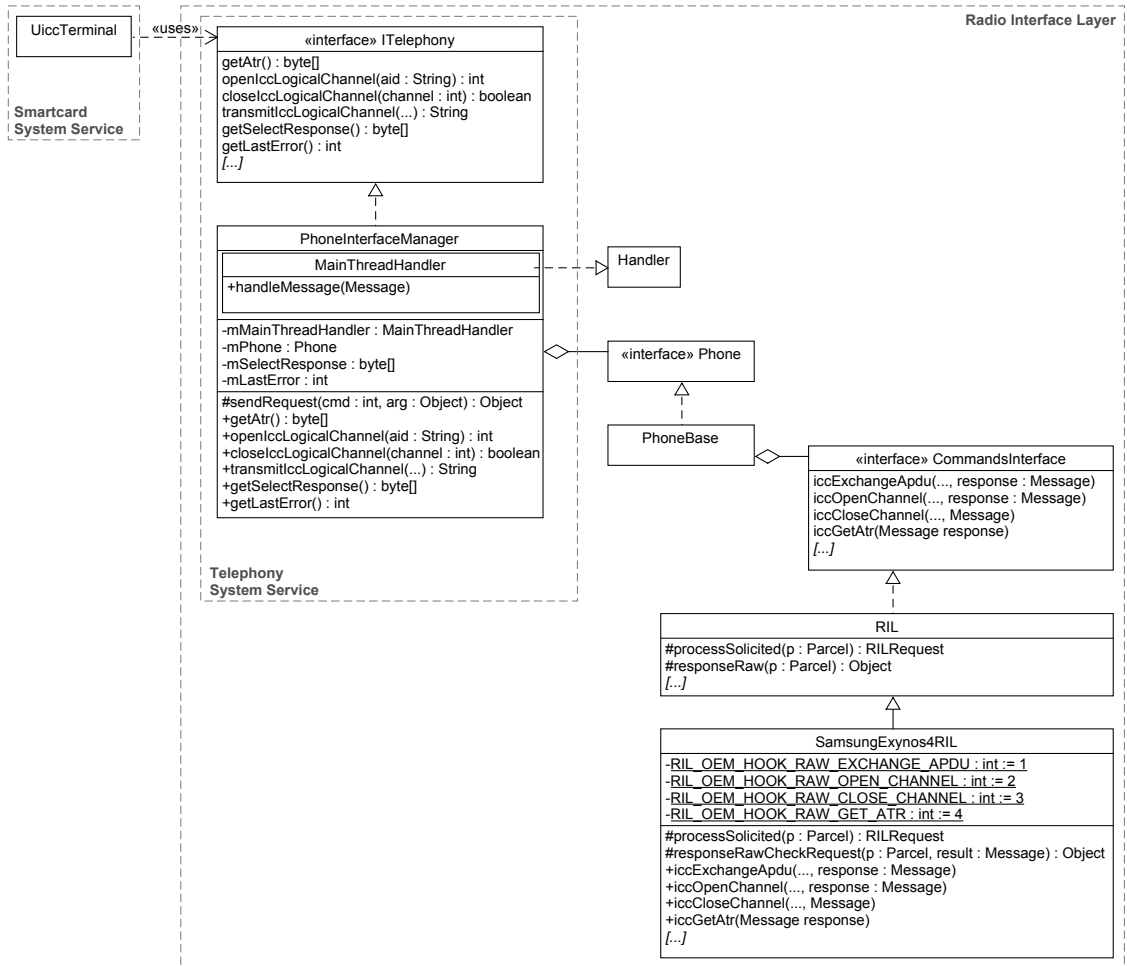

Figure 12: Class diagram for the relevant classes of SEEK and the telephony system service

Next, the above methods are overridden to assemble the command packets for `RIL_REQUEST_OEM_HOOK_RAW` (cf. section 5.4), to add the command tag as parameter to the result message object, and to send the commands through `RIL_REQUEST_OEM_HOOK_RAW`. For instance, the following code shows parts of the implementation of `iccOpenChannel(...)`:

1. First, an output steam, used to generate a byte array that holds the command, is created:

```
@Override
public void iccOpenChannel(String aid, Message result) {
    ByteArrayOutputStream bos = new ByteArrayOutputStream();
    DataOutputStream dos = new DataOutputStream(bos);
```

2. Second, the command code (`0x1509` for opening a logical channel), the length of the command, and the command parameters are written into the stream:

```
    dos.writeByte(0x15);
    dos.writeByte(0x09);
    dos.writeShort(len);

    (...)
```

3. A byte array representation of the custom command is generated from the output stream:

```
    byte[] rawRequest = bos.toByteArray();
```

4. The resulting message is tagged with our custom command code constant (`RIL_OEM_HOOK_RAW_OPEN_CHANNEL`) by storing that value in the (otherwise unused) field `arg1` of the result message object:

```
    result.arg1 = RIL_OEM_HOOK_RAW_OPEN_CHANNEL;
```

5. Finally, an instance of a `RIL_REQUEST_OEM_HOOK_RAW` RIL command is obtained, the byte array representation of the custom command is added, and the request is sent to the RIL system daemon:

```
    RILRequest rr = RILRequest.obtain(
                            RIL_REQUEST_OEM_HOOK_RAW,
                            result);
    rr.mParcel.writeByteArray(rawRequest);
    send(rr);
}
```

After implementing similar processing for the other SEEK-specific methods, it is possible to perform logical channel management and to send smartcard commands to the UICC. As a next step, the responses to these commands need to be detected and decoded. All three commands are wrapped inside a generic `RIL_REQUEST_OEM_HOOK_RAW` RIL request and are additionally tagged with their actual meaning. Therefore, the responses to this generic request must be intercepted, evaluated with regard to their command tag, and decoded into the format expected by SEEK.

Responses to `RIL_REQUEST_OEM_HOOK_RAW` are handled inside the method `process-Solicited()`. By default, responses to this request are passed to the method `responseRaw()` which extracts the response data into a byte array. Therefore, our implementation adds a new method `responseRawCheckRequest()` that performs a conversion based on the actual command (as tagged in the result message) and replaces the original response processing for `RIL_REQUEST_OEM_HOOK_RAW`:

```
@Override
protected RILRequest processSolicited (Parcel p) {

    (...)

    case RIL_REQUEST_RESET_RADIO:
        ret = responseVoid(p);
        break;
    case RIL_REQUEST_OEM_HOOK_RAW:
        ret = responseRawCheckRequest(p, rr.mResult);
        break;

    (...)
```

The method `responseRawCheckRequest()` checks the command tag in `arg1` of the result message object `rr.mResult`. Based on the command tag, it decodes the response bytes into the format expected by the SEEK implementation:

```
protected Object responseRawCheckRequest(Parcel p,
                                         Message result) {
    Object ret = null;
    switch(result.arg1) {
        case RIL_OEM_HOOK_RAW_EXCHANGE_APDU: (...)
        case RIL_OEM_HOOK_RAW_OPEN_CHANNEL:  (...)
        case RIL_OEM_HOOK_RAW_CLOSE_CHANNEL: ret = null; break;
        case RIL_OEM_HOOK_RAW_GET_ATR:       (...)
        default: ret = responseRaw(p); break;
    }
    return ret;
}
```

For each of the commands, the response is obtained by creating a byte array from the response parcel object `p`:

```
byte[] responseRaw = p.createByteArray();
```

For `RIL_OEM_HOOK_RAW_EXCHANGE_APDU`, that byte array is the response APDU itself. As SEEK expects an instance of the class `IccIoResult` as result, the byte array needs to be cut into the data part and the two bytes of the status word to create a new instance of `IccIoResult`:

```java
byte[] data = new byte[responseRaw.length - 2];
System.arraycopy(responseRaw, 0, data, 0, data.length);
ret = new IccIoResult(
        responseRaw[responseRaw.length - 2] & 0x0ff,
        responseRaw[responseRaw.length - 1] & 0x0ff,
        data);
```

For `RIL_OEM_HOOK_RAW_OPEN_CHANNEL`, the byte array contains the channel ID and the SELECT response, each preceded by a length byte. SEEK expects an integer array (`int[]`) where the first array entry contains the channel ID. Therefore, the channel ID is converted into an integer value and stored as the first value of an `int[]` array. In addition, the SELECT response is stored into that same array by casting each byte of the SELECT response into an integer value. These additional array elements are ignored by the default implementation of SEEK in CyanogenMod 11.0.

```java
int idLen = responseRaw[0] & 0x0ff;
int channelId = 0;
for (int i = idLen; i >= 1; --i) {
    channelId <<= 8;
    channelId |= responseRaw[i] & 0x0ff;
}

int selectResLen = responseRaw[idLen + 1] & 0x0ff;
byte[] selectRes = new byte[selectResLen];
System.arraycopy(responseRaw, idLen + 2,
                 selectRes, 0, selectResLen);

int[] intArrayRet = new int[1 + selectRes.length];
intArrayRet[0] = channelId;
for (int i = 0; i < selectRes.length; ++i) {
    intArrayRet[1 + i] = selectRes[i] & 0x0ff;
}
ret = intArrayRet;
```

For `RIL_OEM_HOOK_RAW_GET_ATR`, the byte array contains the ATR preceded by a length byte and a second byte of unknown purpose. SEEK expects the ATR bytes as a string of hexadecimal digits:

```java
byte[] atr = new byte[responseRaw[0] & 0x0ff];
System.arraycopy(responseRaw, 2, atr, 0, atr.length);
ret = IccUtils.bytesToHexString(atr);
```

Finally, the error codes defined for the SEEK-specific RIL errors in CyanogenMod do not match the error code values used by the Exynos4 RIL in the Samsung Galaxy S3. Consequently, these device-specific error codes need to be translated to the values expected by the CyanogenMod implementation. This translation is performed in the method `processSolicited()`. If an error is received in response to `RIL_REQUEST_OEM_HOOK_RAW` for one of the tagged commands, the error code values `INVALID_PARAMETER`, `NO_SUCH_ELEMENT`, and `MISSING_RESOURCE` are converted:

```
if (error != 0) {
    switch (rr.mRequest) {
        (...)
        case RIL_REQUEST_OEM_HOOK_RAW:
            if (rr.mResult.arg1 != 0) {
                switch (error) {
                    case 27: error = INVALID_PARAMETER; break;
                    case 29: error = NO_SUCH_ELEMENT; break;
                    case 30: error = MISSING_RESOURCE; break;
                    default: break;
                }
            }
            break;
    }
    rr.onError(error, ret);
    return rr;
}
```

`INVALID_PARAMETER` errors received in response to the standard RIL command `RIL_REQUEST_SIM_IO`, which is also used by SEEK, are translated in a similar way:

```
        case RIL_REQUEST_SIM_IO:
            if (error == 27) {
                error = INVALID_PARAMETER;
            }
            break;
```

### 6.3.2 Telephony System Service

While the default implementation of SEEK included in CyanogenMod 11.0 retrieves the SELECT response by sending an additional GET RESPONSE APDU (using the `transmitIccLogicalChannel()` method) immediately after opening a logical channel for an applet, the Exynos4 RIL in the Galaxy S3 immediately returns the SELECT response in the response to the command that opens a logical channel. Therefore, this SELECT response needs to be stored by the telephony service and an additional method is necessary to pass the stored value from the telephony system service to the UICC terminal module.

This method can be integrated into the telephony service interface by adding the following lines to the IPC interface definition file `ITelephony.aidl` (in `frameworks/base/telephony/java/com/android/internal/telephony/`):

```
/**
 * Get SELECT response from a previous call to
 * openIccLogicalChannel(AID)
 */
byte[] getSelectResponse();
```

Then, this interface can be implemented by the main entry-point implementation of the telephony system service, `PhoneInterfaceManager` (in `packages/services/Telephony/src/com/android/phone/PhoneInterfaceManager.java`):

```
private byte[] mSelectResponse = null;

public byte[] getSelectResponse() {
    return mSelectResponse;
}
```

Thus, the method `getSelectResponse()` returns the last SELECT response stored in `mSelectResponse`.

The result of the `openIccLogicalChannel()` command is received as the message `EVENT_OPEN_CHANNEL_DONE` within the `MainThreadHandler` in `PhoneInterface-Manager`. This result contains the ID of the opened logical channel as well as the SELECT response, both combined into one integer array. Therefore, the SELECT response needs to be extracted from that integer array:

```
case EVENT_OPEN_CHANNEL_DONE:
    (...)
    int[] resultArray = (int[]) ar.result;
    request.result = new Integer(resultArray[0]);
    mSelectResponse = null;
    if (resultArray.length > 1) {
        mSelectResponse = new byte[resultArray.length - 1];
        for (int i = 1; i < resultArray.length; ++i) {
            mSelectResponse[i - 1] =
                    (byte)(resultArray[i] & 0x0ff);
        }
    }
    mLastError = SUCCESS;
    (...)
```

### 6.3.3 Smartcard System Service

Finally, the UICC terminal of the SEEK implementation needs to be adapted to read the SELECT response by calling the method `getSelectResponse()` in-stead of issuing an additional GET RESPONSE APDU exchange. This is done inside the method `internalOpenLogicalChannel()` of the `UiccTerminal` class of the smartcard system service (in `packages/apps/SmartCardService/src/org/simalliance/openmobileapi/service/terminals/UiccTerminal.java`):

```
mSelectResponse = manager.getSelectResponse();
```

In order to remain compatible with other RIL implementations (e.g. emulator extensions) that do not return the SELECT response that way, the GET RESPONSE APDU is still issued in case `getSelectResponse()` returns `null`:

```java
if (mSelectResponse == null) {
    byte[] getResponseCmd = new byte[] {
                    0x00, (byte) 0xC0, 0x00, 0x00, 0x00 };
    (...)
    mSelectResponse = internalTransmit(getResponseCmd);
}
```

## 6.4 Building CyanogenMod

After applying all the modifications, the last step is to build the system image for the Samsung Galaxy S3 (i9300) and to install the resulting over-the-air update `.zip` file:

```
$ source build/envsetup.sh
$ brunch i9300
```

The update `.zip`, that can be installed on the phone, is created under `out/target/product/i9300/cm-11-YYYYMMDD-UNOFFICIAL-i9300.zip`.

# 7. Summary and Outlook

We gave an overview of secure element integration into mobile devices, the Open Mobile API and its implementation for Android, the SEEK-for-Android project. We found that several current Android devices, particularly devices manufactured by Samsung, ship with an implementation of the Open Mobile API that allows access to the UICC and possibly to other secure elements.

While all these implementations seem to be based on the SEEK-for-Android project, we discovered that access to the UICC is usually facilitated through non-standard, platform-specific interfaces. As a consequence, access to such secure elements is not available in customized Android ROMs even when integrating the open-source implementation of the Open Mobile API provided by the SEEK project.

In order to overcome this limitation, we assembled a toolbox for reverse-engineering stock ROMs of Android devices. We used this toolbox to statically analyze the stock ROM of a Samsung Galaxy S3 (international version) to find out how that existing implementation accesses the UICC through the radio interface. Finally, we reimplemented UICC access for this device in CyanogenMod 11.0.

As we found that CyanogenMod uses one common implementation to access the radio interface library on all Exynos4-based devices (like the Galaxy S3 and the Galaxy S2), we assume that all these devices may also use the same commands to interact with the UICC. Hence, future tests could verify if the same methods are usable on other Exynos4-based Android devices.

Moreover, the reverse-engineering toolbox could be used to analyze the implementations of other devices in order to implement support for access to UICC-based secure elements in CyanogenMod on a broader range of mobile devices.

# References

[1] ECMA-373: Near Field Communication Wired Interface (NFC-WI). Rev. 1 (Jun 2006)

[2] ETSI TS 102 613: Smart Cards; UICC – Contactless Front-end (CLF) Interface; Part 1: Physical and data link layer characteristics (Release 11). Technical specification, V11.0.0 (Sep 2012)

[3] ETSI TS 127 007: Digital cellular telecommunications system (Phase 2+); Universal Mobile Telecommunications System (UMTS); AT command set for User Equipment (UE) (3GPP TS 27.007 version 8.3.0 Release 8). Technical specification, V8.3.0 (Apr 2008)

[4] Giesecke & Devrient: SEEK-for-Android – Secure Element Evaluation Kit for the Android platform, Open Source Project, http://seek-for-android.github.io/

[5] Giesecke & Devrient: Using SmartCard API, SEEK-for-Android, https://github.com/seek-for-android/pool/wiki/UsingSmartCardAPI

[6] Giesecke & Devrient: RIL Extension Specification. V0.2 (Oct 2010)

[7] Giesecke & Devrient: RIL Extension Specification Addendum A. V0.1 (Oct 2010)

[8] GlobalPlatform: Secure Element Access Control. Specification, Version 1.1 (Sep 2014)

[9] ISO/IEC 7816-3: Identification cards – Integrated circuit(s) cards with contacts – Electronic signals and transmission protcols

[10] ISO/IEC 7816-4: Identification cards – Integrated circuit(s) cards with contacts – Interindustry commands for interchange

[11] NFC Forum: NFC Controller Interface (NCI). Technical specification, 1.1 (Jul 2014)

[12] Roland, M.: Security Issues in Mobile NFC Devices. T-Labs Series in Telecommunication Services, Springer (2015)

[13] RSA Laboratories: PKCS #15 v1.1: Cryptographic Token Information Syntax Standard (Jun 2000)

[14] SD Card Association: SD Specifications – Part A1 Advanced Security SD Extension Simplified Specification. Version 2.00 (May 2010)

[15] SD Card Association: SD Specifications – Part 1 Physical Layer Simplified Specification – NFC (Near Field Communication) Interface Simplified Addendum. Version 1.00 (Nov 2013)

[16] SD Card Association: Activating New Mobile Services and Business Models with smartSD Memory cards. White paper, revised version (Nov 2014)

[17] SIMalliance: NFC Secure Element Stepping Stones (Jul 2013)

[18] SIMalliance: Open Mobile API specification, V2.05 (Feb 2014)

# Appendix A. Reverse-Engineering Examples

This section contains examples of disassembled Dalvik code and its automated translation into Java source code using our reverse-engineering toolchain.

## A.1 Method IccUtils.bytesToHexString()

### A.1.1 Smali Assembler

```
1   .method public static bytesToHexString([B)Ljava/lang/String;
2     .registers 5
3     .param p0, "bytes"    # [B
4
5     .prologue
6     .line 396
7     if-nez p0, :cond_4
8
9     const/4 v3, 0x0
10
11    .line 412
12    :goto_3
13    return-object v3
14
15    .line 398
16    :cond_4
17    new-instance v2, Ljava/lang/StringBuilder;
18    array-length v3, p0
19    mul-int/lit8 v3, v3, 0x2
20    invoke-direct {v2, v3}, Ljava/lang/StringBuilder;-><init>(I)V
21
22    .line 400
23    .local v2, "ret":Ljava/lang/StringBuilder;
24    const/4 v1, 0x0
25
26    .local v1, "i":I
27    :goto_d
28    array-length v3, p0
29    if-ge v1, v3, :cond_2f
30
31    .line 403
32    aget-byte v3, p0, v1
33    shr-int/lit8 v3, v3, 0x4
34    and-int/lit8 v0, v3, 0xf
35
36    .line 405
37    .local v0, "b":I
38    const-string v3, "0123456789abcdef"
39    invoke-virtual {v3, v0}, Ljava/lang/String;->charAt(I)C
40    move-result v3
```

```
41    invoke-virtual {v2, v3}, Ljava/lang/StringBuilder;->append(C)
          Ljava/lang/StringBuilder;
42
43    .line 407
44    aget-byte v3, p0, v1
45    and-int/lit8 v0, v3, 0xf
46
47    .line 409
48    const-string v3, "0123456789abcdef"
49    invoke-virtual {v3, v0}, Ljava/lang/String;->charAt(I)C
50    move-result v3
51    invoke-virtual {v2, v3}, Ljava/lang/StringBuilder;->append(C)
          Ljava/lang/StringBuilder;
52
53    .line 400
54    add-int/lit8 v1, v1, 0x1
55    goto :goto_d
56
57    .line 412
58    .end local v0    # "b":I
59    :cond_2f
60    invoke-virtual {v2}, Ljava/lang/StringBuilder;->toString()
          Ljava/lang/String;
61    move-result-object v3
62    goto :goto_3
63 .end method
```

### A.1.2  Generated Java Source Code

```java
1  public static String bytesToHexString(byte[] paramArrayOfByte) {
2      if (paramArrayOfByte == null) {
3          return null;
4      }
5      StringBuilder localStringBuilder =
6                  new StringBuilder(paramArrayOfByte.length * 2);
7      int i = 0;
8      while (i < paramArrayOfByte.length) {
9          localStringBuilder.append("0123456789abcdef"
10                 .charAt(paramArrayOfByte[i] >> 4 & 0xF));
11         localStringBuilder.append("0123456789abcdef"
12                 .charAt(paramArrayOfByte[i] & 0xF));
13         i += 1;
14     }
15     return localStringBuilder.toString();
16 }
```

### A.1.3  Java Source Code Enriched with Information from Disassembly

```java
1  public static String bytesToHexString(byte[] bytes) {
```

```java
2          if (bytes == null) {
3              return null;
4          }
5          StringBuilder ret = new StringBuilder(bytes.length * 2);
6          for (int i = 0; i < bytes.length; ++i) {
7              int b = (bytes[i] >> 4) & 0xF;
8              ret.append("0123456789abcdef".charAt(b));
9              b = bytes[i] & 0xF;
10             ret.append("0123456789abcdef".charAt(b));
11         }
12         return ret.toString();
13  }
```

## A.2  Method PhoneInterfaceManager.openIccLogicalChannel()

### A.2.1  Smali Assembler

```smali
1   .method public openIccLogicalChannel(Ljava/lang/String;)I
2     .registers 6
3     .param p1, "AID"    # Ljava/lang/String;
4
5     .prologue
6     .line 922
7     sget-boolean v1, Lcom/android/phone/PhoneInterfaceManager;->DBG_ENG:Z
8     if-eqz v1, :cond_1c
9
10    const-string v1, "PhoneInterfaceManager"
11
12    new-instance v2, Ljava/lang/StringBuilder;
13    invoke-direct {v2}, Ljava/lang/StringBuilder;-><init>()V
14
15    const-string v3, "> openIccLogicalChannel "
16    invoke-virtual {v2, v3}, Ljava/lang/StringBuilder;->append(Ljava/lang/String
        ;)Ljava/lang/StringBuilder;
17    move-result-object v2
18
19    invoke-virtual {v2, p1}, Ljava/lang/StringBuilder;->append(Ljava/lang/String
        ;)Ljava/lang/StringBuilder;
20    move-result-object v2
21
22    invoke-virtual {v2}, Ljava/lang/StringBuilder;->toString()Ljava/lang/String;
23    move-result-object v2
24
25    invoke-static {v1, v2}, Lcom/android/phone/Log;->d(Ljava/lang/String;
        Ljava/lang/String;)I
26
27    .line 923
28    :cond_1c
29    const/16 v1, 0xe
```

```
30
31    new-instance v2, Lcom/android/phone/PhoneInterfaceManager$IccOpenChannel;
32    invoke-direct {v2, p1},
          Lcom/android/phone/PhoneInterfaceManager$IccOpenChannel;-><init>(
          Ljava/lang/String;)V
33
34    invoke-direct {p0, v1, v2}, Lcom/android/phone/PhoneInterfaceManager;->
          sendRequest(ILjava/lang/Object;)Ljava/lang/Object;
35    move-result-object v0
36    check-cast v0, Ljava/lang/Integer;
37
38    .line 925
39    .local v0, "channel":Ljava/lang/Integer;
40    sget-boolean v1, Lcom/android/phone/PhoneInterfaceManager;->DBG_ENG:Z
41    if-eqz v1, :cond_45
42
43    const-string v1, "PhoneInterfaceManager"
44
45    new-instance v2, Ljava/lang/StringBuilder;
46    invoke-direct {v2}, Ljava/lang/StringBuilder;-><init>()V
47
48    const-string v3, "<␣openIccLogicalChannel␣"
49    invoke-virtual {v2, v3}, Ljava/lang/StringBuilder;->append(Ljava/lang/String
          ;)Ljava/lang/StringBuilder;
50    move-result-object v2
51
52    invoke-virtual {v2, v0}, Ljava/lang/StringBuilder;->append(Ljava/lang/Object
          ;)Ljava/lang/StringBuilder;
53    move-result-object v2
54
55    invoke-virtual {v2}, Ljava/lang/StringBuilder;->toString()Ljava/lang/String;
56    move-result-object v2
57
58    invoke-static {v1, v2}, Lcom/android/phone/Log;->d(Ljava/lang/String;
          Ljava/lang/String;)I
59
60    .line 926
61    :cond_45
62    invoke-virtual {v0}, Ljava/lang/Integer;->intValue()I
63    move-result v1
64    return v1
65  .end method
```

### A.2.2  Generated Java Source Code

```java
1  public int openIccLogicalChannel(String paramString) {
2      if (DBG_ENG) {
3          Log.d("PhoneInterfaceManager",
4                   ">␣openIccLogicalChannel␣" + paramString);
5      }
6      paramString = (Integer)sendRequest(14,
```

```
 7                                        new IccOpenChannel(paramString));
 8      if (DBG_ENG) {
 9          Log.d("PhoneInterfaceManager",
10              "<␣openIccLogicalChannel␣" + paramString);
11      }
12      return paramString.intValue();
13  }
```

### A.2.3  Java Source Code Enriched with Information from Disassembly

```
 1  public int openIccLogicalChannel(String AID) {
 2      if (DBG_ENG) {
 3          Log.d("PhoneInterfaceManager",
 4              ">␣openIccLogicalChannel␣" + AID);
 5      }
 6      Integer channel = (Integer)sendRequest(14,
 7                                        new IccOpenChannel(AID));
 8      if (DBG_ENG) {
 9          Log.d("PhoneInterfaceManager",
10              "<␣openIccLogicalChannel␣" + channel);
11      }
12      return channel.intValue();
13  }
```

## A.3  Method IccIoResult.getException()

### A.3.1  Smali Assembler

```
 1  .method public getException()Lcom/android/internal/telephony/IccException;
 2      .registers 4
 3
 4      .prologue
 5      .line 57
 6      invoke-virtual {p0}, Lcom/android/internal/telephony/IccIoResult;->success()
              Z
 7      move-result v0
 8      if-eqz v0, :cond_8
 9
10      const/4 v0, 0x0
11
12      .line 67
13      :goto_7
14      return-object v0
15
16      .line 59
17      :cond_8
18      iget v0, p0, Lcom/android/internal/telephony/IccIoResult;->sw1:I
19      packed-switch v0, :pswitch_data_48
```

```
20
21      .line 67
22      new-instance v0, Lcom/android/internal/telephony/IccException;
23
24      new-instance v1, Ljava/lang/StringBuilder;
25      invoke-direct {v1}, Ljava/lang/StringBuilder;-><init>()V
26
27      const-string/jumbo v2, "sw1:"
28      invoke-virtual {v1, v2}, Ljava/lang/StringBuilder;->append(Ljava/lang/String
            ;)Ljava/lang/StringBuilder;
29      move-result-object v1
30
31      iget v2, p0, Lcom/android/internal/telephony/IccIoResult;->sw1:I
32      invoke-virtual {v1, v2}, Ljava/lang/StringBuilder;->append(I)
            Ljava/lang/StringBuilder;
33      move-result-object v1
34
35      const-string v2, "␣sw2:"
36      invoke-virtual {v1, v2}, Ljava/lang/StringBuilder;->append(Ljava/lang/String
            ;)Ljava/lang/StringBuilder;
37      move-result-object v1
38
39      iget v2, p0, Lcom/android/internal/telephony/IccIoResult;->sw2:I
40      invoke-virtual {v1, v2}, Ljava/lang/StringBuilder;->append(I)
            Ljava/lang/StringBuilder;
41      move-result-object v1
42
43      invoke-virtual {v1}, Ljava/lang/StringBuilder;->toString()Ljava/lang/String;
44      move-result-object v1
45
46      invoke-direct {v0, v1}, Lcom/android/internal/telephony/IccException;-><init
            >(Ljava/lang/String;)V
47      goto :goto_7
48
49      .line 61
50      :pswitch_35
51      iget v0, p0, Lcom/android/internal/telephony/IccIoResult;->sw2:I
52      const/16 v1, 0x8
53      if-ne v0, v1, :cond_41
54
55      .line 62
56      new-instance v0, Lcom/android/internal/telephony/IccFileTypeMismatch;
57      invoke-direct {v0}, Lcom/android/internal/telephony/IccFileTypeMismatch;-><
            init>()V
58      goto :goto_7
59
60      .line 64
61      :cond_41
62      new-instance v0, Lcom/android/internal/telephony/IccFileNotFound;
63      invoke-direct {v0}, Lcom/android/internal/telephony/IccFileNotFound;-><init
            >()V
64      goto :goto_7
```

```
65
66    .line 59
67    nop
68
69    :pswitch_data_48
70    .packed-switch 0x94
71        :pswitch_35
72    .end packed-switch
73 .end method
```

### A.3.2  Generated Java Source Code

```java
1  public IccException getException() {
2      if (success()) {
3          return null;
4      }
5      switch (this.sw1) {
6          default:
7              return new IccException("sw1:" + this.sw1 +
8                                      "␣sw2:" + this.sw2);
9      }
10     if (this.sw2 == 8) {
11         return new IccFileTypeMismatch();
12     }
13     return new IccFileNotFound();
14 }
```

### A.3.3  Java Source Code Enriched with Information from Disassembly

```java
1  public IccException getException() {
2      if (success()) {
3          return null;
4      }
5      switch (this.sw1) {
6          case 0x94:
7              if (this.sw2 == 8) {
8                  return new IccFileTypeMismatch();
9              }
10             return new IccFileNotFound();
11
12         default:
13             return new IccException("sw1:" + this.sw1 +
14                                     "␣sw2:" + this.sw2);
15     }
16 }
```

# Appendix B.  Implementation: Telephony System Service

This appendix lists the reverse-engineered implementation of the UICC-access functionality of the telephony system service and relevant framework classes. Some optimizations were added to the reverse-engineered Java source code. Moreover, debug information (like output of log messages) was removed.

## B.1  Class PhoneInterfaceManager

This class is part of the application package /system/app/SecPhone.*.

```
1   package com.android.phone;
    (...)
3   import android.os.AsyncResult;
4   import android.os.Handler;
5   import android.os.Looper;
6   import android.os.Message;
7   import com.android.internal.telephony.ITelephony.Stub;
8   import com.android.internal.telephony.IccIoResult;
9   import com.android.internal.telephony.IccUtils;
10  import com.android.internal.telephony.Phone;
11  import java.io.ByteArrayOutputStream;
12  import java.io.DataOutputStream;
13  import java.io.IOException;
    (...)
15  public class PhoneInterfaceManager extends ITelephony.Stub {
    (...)
17      private static final int CMD_EXCHANGE_APDU = 12;
18      private static final int EVENT_EXCHANGE_APDU_DONE = 13;
19      private static final int CMD_OPEN_CHANNEL = 14;
20      private static final int EVENT_OPEN_CHANNEL_DONE = 15;
21      private static final int CMD_CLOSE_CHANNEL = 16;
22      private static final int EVENT_CLOSE_CHANNEL_DONE = 17;
23      private static final int CMD_GET_ATR_INFO = 18;
24      private static final int EVENT_GET_ATR_INFO_DONE = 19;
25

        (...)
27      private int lastError;
28      MainThreadHandler mMainThreadHandler;
29      Phone mPhone;
30      byte[] mSelectResponse = null;
        (...)
```

```
32      private String exchangeIccAPDU(int cla, int ins,
33                                      int channelId,
34                                      int p1, int p2,
35                                      int len, String dataStr) {
36          IccIoResult result = (IccIoResult)sendRequest(
37                  CMD_EXCHANGE_APDU,
38                  new IccAPDUArgument(cla, ins,
39                                      channelId,
40                                      p1, p2,
41                                      len, dataStr));
42
43          String sw = Integer.toHexString((result.sw1 << 8) +
44                                          result.sw2 +
45                                          0x10000).substring(1);
46          String respApdu = sw;
47          if (result.payload != null) {
48              respApdu = IccUtils.bytesToHexString(result.payload) + sw;
49          }
50          return respApdu;
51      }
        (...)
53      private Object sendRequest(int command, Object argument) {
54          if (Looper.myLooper() == mMainThreadHandler.getLooper()) {
55              throw new RuntimeException("This method will deadlock if
                    called from the main thread.");
56          }
57
58          MainThreadRequest request = new MainThreadRequest(argument);
59          Message msg = mMainThreadHandler.obtainMessage(command,
60                                                         request);
61          msg.sendToTarget();
62
63          synchronized (request) {
64              while (request.result == null) {
65                  try {
66                      request.wait();
67                  } catch (InterruptedException e) {}
68              }
69          }
70
71          return request.result;
72      }
        (...)
74      public boolean closeIccLogicalChannel(int channelId) {
```

```
75          Integer result = (Integer)sendRequest(
76                  CMD_CLOSE_CHANNEL, new IccCloseChannel(channelId));
77          return result.intValue() != -1;
78      }
        (...)
80      public byte[] getAtr() {
81          return (byte[])sendRequest(CMD_GET_ATR_INFO, null);
82      }
        (...)
84      public int getLastError() {
85          return lastError;
86      }
        (...)
88      public byte[] getSelectResponse() {
89          return mSelectResponse;
90      }
        (...)
92      public int openIccLogicalChannel(String aidStr) {
93          Integer result = (Integer)sendRequest(
94                  CMD_OPEN_CHANNEL, new IccOpenChannel(aidStr));
95          return result.intValue();
96      }
        (...)
98      public String transmitIccLogicalChannel(int cla, int ins,
99                                              int channelId,
100                                             int p1, int p2,
101                                             int len, String dataStr) {
102         return exchangeIccAPDU(cla, ins, channelId,
103                         p1, p2, len, dataStr);
104     }
        (...)
106     private static final class IccAPDUArgument {
107         public int channel;
108         public int cla;
109         public int command;
110         public String data;
111         public int p1;
112         public int p2;
113         public int p3;
114
115         public IccAPDUArgument(int cla, int ins,
116                         int channelId,
```

```
117                                     int p1, int p2,
118                                     int len, String dataStr) {
119                 this.channel = channelId;
120                 this.cla = cla;
121                 this.command = ins;
122                 this.p1 = p1;
123                 this.p2 = p2;
124                 this.p3 = len;
125                 this.data = dataStr;
126             }
127         }
128
129     private static final class IccCloseChannel {
130         public int sessionID;
131
132         public IccCloseChannel(int channelId) {
133             this.sessionID = channelId;
134         }
135     }
136
137     private static final class IccOpenChannel {
138         public String AID;
139
140         public IccOpenChannel(String aidStr) {
141             this.AID = aidStr;
142         }
143     }
        (...)
145     private final class MainThreadHandler extends Handler {
146         private MainThreadHandler() {}
147
148         @Override
149         public void handleMessage(Message msg) {
150             MainThreadRequest request;
151             AsyncResult ar;
152             byte[] data;
153
154             switch (msg.what) {
            (...)
156                 break;
157
158             case CMD_EXCHANGE_APDU:
159                 request = (MainThreadRequest)msg.obj;
160                 IccAPDUArgument argumentAPDU =
```

```java
161                         (IccAPDUArgument)request.argument;
162
163             ByteArrayOutputStream bos =
164                     new ByteArrayOutputStream();
165             DataOutputStream dos =
166                     new DataOutputStream(bos);
167
168             int len = 9;
169
170             if (argumentAPDU.data != null) {
171                 len += argumentAPDU.data.length() / 2;
172             }
173
174             if (argumentAPDU.p3 == -1) {
175                 --len;
176             }
177
178             try {
179                 dos.writeByte(0x15);
180                 if (argumentAPDU.channel == 0) {
181                     dos.writeByte(0x08);
182                     dos.writeShort(len);
183                 } else {
184                     if (argumentAPDU.p3 != -1) {
185                         dos.writeByte(0x0B);
186                     } else {
187                         dos.writeByte(0x0C);
188                     }
189                     dos.writeShort(len + 4);
190                 }
191                 dos.writeByte(argumentAPDU.cla);
192                 dos.writeByte(argumentAPDU.command);
193                 dos.writeByte(argumentAPDU.p1);
194                 dos.writeByte(argumentAPDU.p2);
195                 if (argumentAPDU.p3 != -1) {
196                     dos.writeByte(argumentAPDU.p3);
197                 }
198                 if (argumentAPDU.channel != 0) {
199                     dos.writeInt(argumentAPDU.channel);
200                 }
201                 if (argumentAPDU.data != null) {
202                     byte[] ba = new byte[
203                             argumentAPDU.data.length() / 2];
204                     for (int i = 0; i < ba.length; ++i) {
205                         ba[i] = (byte)Integer.parseInt(
206                                 argumentAPDU.data.substring(
```

```
207                                             i * 2, i * 2 + 2),
208                                     16);
209                         }
210                         dos.write(ba);
211                     }
212                 } catch (IOException e) {}
213
214             mPhone.invokeOemRilRequestRaw(
215                     bos.toByteArray(),
216                     obtainMessage(EVENT_EXCHANGE_APDU_DONE,
217                             request));
218
219             try {
220                 dos.close();
221             } catch (IOException e) {}
222             break;
223
224         case EVENT_EXCHANGE_APDU_DONE:
225             ar = (AsyncResult)msg.obj;
226             request = (MainThreadRequest)ar.userObj;
227
228             if ((ar.exception == null) && (ar.result != null) &&
229                 (((byte[])ar.result).length >= 2)) {
230                 byte[] b = (byte[])ar.result;
231
232                 if (b.length > 2) {
233                     data = new byte[b.length - 2];
234                     System.arraycopy(b, 0, data, 0, data.length);
235                 } else {
236                     data = null;
237                 }
238
239                 request.result = new IccIoResult(
240                         b[b.length - 2] & 0x0FF,
241                         b[b.length - 1] & 0x0FF,
242                         data);
243                 lastError = 0;
244             } else {
245                 request.result = new IccIoResult(0x6F, 0x00,
246                                                 null);
247                 lastError = 1;
248
249                 if ((ar.exception != null) &&
250                     (ar.exception instanceof CommandException)) {
251                     CommandException e =
252                             (CommandException)ar.exception;
```

```
253                         if (e.getCommandError() ==
254                             CommandException.Error.INVALID_PARAMETER) {
255                             lastError = 5;
256                         }
257                     }
258                 }
259
260             synchronized (request) {
261                 request.notifyAll();
262             }
263             break;
264
265         case CMD_OPEN_CHANNEL:
266             request = (MainThreadRequest)msg.obj;
267             IccOpenChannel openArgument =
268                     (IccOpenChannel)request.argument;
269
270             ByteArrayOutputStream bos_open =
271                     new ByteArrayOutputStream();
272             DataOutputStream dos_open =
273                     new DataOutputStream(bos_open);
274
275             int len_open = 4;
276
277             if (openArgument.AID != null) {
278                 len_open += openArgument.AID.length() / 2;
279             }
280
281             try {
282                 dos_open.writeByte(0x15);
283                 dos_open.writeByte(0x09);
284                 dos_open.writeShort(len_open);
285                 if (openArgument.AID != null) {
286                     byte[] ba = new byte[
287                             openArgument.AID.length() / 2];
288                     for (int i = 0; i < ba.length; ++i) {
289                         ba[i] = (byte)Integer.parseInt(
290                                     openArgument.AID.substring(
291                                         i * 2, i * 2 + 2),
292                                     16);
293                     }
294                     dos_open.write(ba);
295                 }
296             } catch (IOException e) {}
297
298             mPhone.invokeOemRilRequestRaw(
```

```
299                             bos_open.toByteArray(),
300                             obtainMessage(EVENT_OPEN_CHANNEL_DONE,
301                                     request));
302
303                 try {
304                     dos_open.close();
305                 } catch (IOException e) {}
306                 break;
307
308             case EVENT_OPEN_CHANNEL_DONE:
309                 ar = (AsyncResult)msg.obj;
310                 request = (MainThreadRequest)ar.userObj;
311
312                 if ((ar.exception == null) && (ar.result != null)) {
313                     data = (byte[])ar.result;
314                     int id_len = data[0];
315                     int select_res_len = data[id_len + 1];
316
317                     int channelId = 0;
318                     for (int i = id_len; i >= 1; --i) {
319                         channelId <<= 8;
320                         channelId |= data[i] & 0x0FF;
321                     }
322
323                     if (select_res_len > 0) {
324                         mSelectResponse = new byte[select_res_len];
325                         System.arraycopy(data, id_len + 2,
326                                     mSelectResponse, 0,
327                                     select_res_len);
328                     } else {
329                         mSelectResponse = null;
330                     }
331
332                     request.result = new Integer(channelId);
333                     lastError = 0;
334                 } else {
335                     request.result = new Integer(0);
336                     lastError = 1;
337
338                     if ((ar.exception != null) &&
339                         (ar.exception instanceof CommandException)) {
340                         CommandException e =
341                                 (CommandException)ar.exception;
342                         if (e.getCommandError() ==
343                             CommandException.Error.MISSING_RESOURCE) {
344                             lastError = 2;
```

```
345                         } else if (e.getCommandError() ==
346                             CommandException.Error.NO_SUCH_ELEMENT) {
347                             lastError = 3;
348                         }
349                     }
350                 }
351
352                 synchronized (request) {
353                     request.notifyAll();
354                 }
355                 break;
356
357             case CMD_CLOSE_CHANNEL:
358                 request = (MainThreadRequest)msg.obj;
359                 IccCloseChannel closeArgument =
360                         (IccCloseChannel)request.argument;
361
362                 ByteArrayOutputStream bos_close =
363                         new ByteArrayOutputStream();
364                 DataOutputStream dos_close =
365                         new DataOutputStream(bos_close);
366
367                 int len_close = 4;
368
369                 if (closeArgument.sessionID != 0) {
370                     len_close += 4;
371                 }
372
373                 try {
374                     dos_close.writeByte(0x15);
375                     dos_close.writeByte(0x0A);
376                     dos_close.writeShort(len_close);
377                     if (closeArgument.sessionID != 0) {
378                         dos_close.writeInt(closeArgument.sessionID);
379                     }
380                 } catch (IOException e) {}
381
382                 mPhone.invokeOemRilRequestRaw(
383                         bos_close.toByteArray(),
384                         obtainMessage(EVENT_CLOSE_CHANNEL_DONE,
385                             request));
386
387                 try {
388                     dos_close.close();
389                 } catch (IOException e) {}
390                 break;
```

```
391
392             case EVENT_CLOSE_CHANNEL_DONE:
393                 ar = (AsyncResult)msg.obj;
394                 request = (MainThreadRequest)ar.userObj;
395
396                 if (ar.exception == null) {
397                     request.result = new Integer(0);
398                     lastError = 0;
399                 } else {
400                     request.result = new Integer(-1);
401                     lastError = 1;
402
403                     if ((ar.exception instanceof CommandException)) {
404                         CommandException e =
405                             (CommandException)ar.exception;
406                         if (e.getCommandError() ==
407                             CommandException.Error.INVALID_PARAMETER) {
408                             lastError = 5;
409                         }
410                     }
411                 }
412
413                 synchronized (request) {
414                     request.notifyAll();
415                 }
416                 break;
417
418             case CMD_GET_ATR_INFO:
419                 request = (MainThreadRequest)msg.obj;
420
421                 ByteArrayOutputStream bos1 =
422                         new ByteArrayOutputStream();
423                 DataOutputStream dos1 =
424                         new DataOutputStream(bos1);
425
426                 try {
427                     dos1.writeByte(0x15);
428                     dos1.writeByte(0x0D);
429                     dos1.writeShort(4);
430                 } catch (IOException e) {}
431
432                 mPhone.invokeOemRilRequestRaw(
433                         bos1.toByteArray(),
434                         obtainMessage(EVENT_GET_ATR_INFO_DONE,
435                             request));
436
```

```
437                    try {
438                        dos1.close();
439                    } catch (IOException e) {}
440                    break;
441
442                case EVENT_GET_ATR_INFO_DONE:
443                    ar = (AsyncResult)msg.obj;
444                    request = (MainThreadRequest)ar.userObj;
445
446                    data = null;
447
448                    if ((ar.exception == null) && (ar.result != null)) {
449                        byte[] result = (byte[])ar.result;
450                        if (result[0] > 0) {
451                            data = new byte[result[0] & 0x0FF];
452                            System.arraycopy(result, 2,
453                                             data, 0, data.length);
454                        }
455                    }
456                    request.result = data;
457
458                    synchronized (request) {
459                        request.notifyAll();
460                    }
461                    break;
               (...)
463                default:
464                    return;
465                }
466            }
467        }
468
469    private static final class MainThreadRequest {
470        public Object argument;
471        public Object result;
472
473        public MainThreadRequest(Object argument) {
474            this.argument = argument;
475        }
476    }
       (...)
478 }
```

## B.2 Class IccIoResult

This class is part of the framework package `/system/framework/framework.odex`.

```
1   package com.android.internal.telephony;
2
3   public class IccIoResult {
4       public byte[] payload;
5       public int sw1;
6       public int sw2;
7
8       public IccIoResult(int sw1, int sw2, String dataStr) {
9           this(sw1, sw2, IccUtils.hexStringToBytes(dataStr));
10      }
11
12      public IccIoResult(int sw1, int sw2, byte[] data) {
13          this.sw1 = sw1;
14          this.sw2 = sw2;
15          this.payload = data;
16      }
17
18      public IccException getException() {
19          if (success()) {
20              return null;
21          }
22          switch (sw1) {
23              case 0x94:
24                  if (sw2 == 0x08) {
25                      return new IccFileTypeMismatch();
26                  } else {
27                      return new IccFileNotFound();
28                  }
29
30              default:
31                  return new IccException("sw1:" + sw1 +
32                                          " sw2:" + sw2);
33          }
34      }
35
36      public boolean success() {
37          return (sw1 == 0x90) || (sw1 == 0x91) ||
38                  (sw1 == 0x9E) || (sw1 == 0x9F);
39      }
40
41      public String toString() {
42          return "IccIoResponse sw1:0x" + Integer.toHexString(sw1)
```

```
43                         + "␣sw2:0x" + Integer.toHexString(sw2);
44     }
45 }
```

## B.3  Class IccUtils

This class is part of the framework package /system/framework/framework.odex.

```
1 package com.android.internal.telephony;
  (. . .)
3 public class IccUtils {
4     private static final String HEX = "0123456789abcdef";
      (. . .)
6     public static String byteToHexString(byte b) {
7         StringBuilder sb = new StringBuilder(2);
8         sb.append(HEX.charAt((b >> 4) & 0xF));
9         sb.append(HEX.charAt(b & 0xF));
10        return sb.toString();
11    }
12
13    public static String bytesToHexString(byte[] bArray) {
14        if (bArray == null) {
15            return null;
16        }
17
18        StringBuilder sb = new StringBuilder(bArray.length * 2);
19        for (int i = 0; i < bArray.length; ++i) {
20            sb.append(HEX.charAt((bArray[i] >> 4) & 0xF));
21            sb.append(HEX.charAt(bArray[i] & 0xF));
22        }
23
24        return sb.toString();
25    }
      (. . .)
27    static int hexCharToInt(char c) {
28        if ((c >= '0') && (c <= '9')) {
29            return c - '0';
30        } else if ((c >= 'A') && (c <= 'F')) {
31            return c - 'A' + 10;
32        } else if ((c >= 'a') && (c <= 'f')) {
33            return c - 'a' + 10;
34        } else {
35            throw new RuntimeException("invalid␣hex␣char␣'" +
```

```
36                                                      c + "'");
37          }
38      }
39
40      public static byte[] hexStringToBytes(String str) {
41          if (str == null) {
42              return null;
43          }
44
45          final int j = str.length();
46          byte[] bArray = new byte[j / 2];
47
48          int i = 0;
49          for (int i = 0; i < j; i += 2) {
50              bArray[i / 2] =
51                      (byte)(hexCharToInt(str.charAt(i)) << 4 |
52                              hexCharToInt(str.charAt(i + 1)));
53          }
54
55          return bArray;
56      }
        (...)
58  }
```